\documentclass[secnumarabic,amssymb, nobibnotes, aps, prd,onecolumn]{revtex4-2}
\usepackage{graphicx,epsfig,youngtab}

\pdfoutput=1
\usepackage{xcolor}
\usepackage{bbold}
\usepackage{amsmath}
\usepackage{hyperref}
\usepackage{amsfonts}
\expandafter\let\csname equation*\endcsname\relax 
\expandafter\let\csname endequation*\endcsname\relax 
\usepackage{bm}
\usepackage{lmodern}
\usepackage{natbib}

\def\x {\bm{x}}
\def\r {\bm{r}}
\def\d {\rm{d}}

\def\k {\bm{k}}
\def\x {\bm{x}}

\newcommand{\D} {{\tilde{\rm{D}}}}
\newcommand{\p}{_{{\text{\tiny$\|$}}}}% ||

\newcommand{\n}{{\bf \hat{n}}}
\newcommand{\HH}{\mathcal{H}\,}
\newcommand{\Hsq}{\mathcal{H}^{(2)}\,}
\newcommand{\Hcub}{\mathcal{H}^{(3)}\,}
\newcommand{\two}{^{\text{\tiny \color{green}{({{2}})}}}}

\newcommand{\<}{\langle}
\renewcommand{\>}{\rangle}

\def\jnlref#1{{\rm#1}}

\def\mnras{\jnlref{MNRAS}}
\def\aapr{\jnlref{A\&A~Rev.}}
\def\pasp{\jnlref{PASP}}
\def\aap{\jnlref{A\&A}}

\newcommand{\averageA}[1]{\left\langle #1 \right\rangle_{\Omega}}

\begin{document}
   % \title{\boldmath Fitting the FLRW spacetime to observation on ultra-small redshifts \\  and the infered Hubble rates: Paper II} 
        % \title{\boldmath Cosmological fitting problem and the Hubble tension: Paper	 II} 
                     \title{\boldmath Consequences of using a smooth cosmic distance in a lumpy Universe: I}
       % \title{\boldmath The emergence of a smooth cosmic distance laAlcock-Paczy\'{n}skidder from a lumpy universe: I}
        % \title{\boldmath Inferred cosmological parameters from fitting to an FLRW  spacetime}
\author{Obinna Umeh }%$^{1,2,3}$
\affiliation {${}^{1}$Institute of Cosmology \& Gravitation, University of Portsmouth, Portsmouth PO1 3FX, United Kingdom
\\
${}^{2}$Department of Physics, University of the Western Cape,
Cape Town 7535, South Africa 
\\
${}^{3}$Department of Mathematics and Applied Mathematics, University of Cape Town, Rondebosch 7701, South Africa}

\email{
obinna.umeh@port.ac.uk
}
\date{\today}

\begin{abstract}
How do we appropriately fit a model based on an idealised Friedmann-Lema\^{i}tre  Robertson-Walker  spacetime to observations made from a single location in a lumpy Universe? We address this question for surveys that measure the imprints of the baryon acoustic oscillation  in galaxy distribution and the  peak apparent magnitude of the Type Ia supernova. 
These observables are related to the cosmological model through the Alcock-Paczy\'{n}ski parameters and the distance-redshift relation. Using the corresponding inhomogeneous spacetime expressions of these as observed data, we perform a parameter inference assuming that the background Friedmann-Lema\^{i}tre  Robertson-Walker model is the correct model of the Universe.  This process allows us to estimate the best fit Hubble rate and the deceleration parameter. 
We find that the inferred Hubble rate from the monopole of the Alcock-Paczy\'{n}ski parameters is in tension with the Hubble rate determined using the distance-redshift relation. The latter gives the best fit Hubble rate for the cosmological expansion. The constraint on the Hubble rate from the Alcock-Paczy\'{n}ski parameters is contaminated by the environment. When the environmental contribution is restricted to modes in the Hubble flow, we find about (9-12)\% discrepancy in the Hubble rate. Finally, we comment on the insufficiency of the method of cosmography in constraining the deceleration parameter.
\end{abstract}

\maketitle
 \setcounter{footnote}{0}
\DeclareGraphicsRule{.wmf}{bmp}{jpg}{}{}
\maketitle

%\tableofcontents
%\maketitle
%\newpage

\section{Introduction}

The distribution of large scale structures in the Universe on ultra-large scales appear homogeneous and isotropic~\cite{Martinez:1998yp,Hogg:2004vw,Ntelis:2018ctq}. On scales of superclusters, there are inhomogeneities~\cite{Boehringer:2021mix}. The Universe seen on scales of  clusters appear very lumpy with sources moving with discernible peculiar velocities within clusters and clusters interacting gravitationally with each other~\cite{Nastas:2011AandA...532L...6N,Chadayammuri:2021tcp}. 
Ellis and Stoeger pointed out in \cite{Ellis:1987zz} that fitting  models based on homogeneous and isotropic  Friedmann-Lema\^{i}tre  Robertson-Walker (FLRW) spacetime to observation of a lumpy Universe leads to a geometry that ignores on average the details present on small scales. 
Such a description of physical reality by the best-fit FLRW model embodies a smoothing scale that is usually not made explicit. 
 Ellis and Stoeger further argued that any relevant averaging procedure that yields a best-fit FLRW model must be performed on the past light-cone of the observer~\cite{EllSto87,Clarkson:2011zq}. This is particularly important because averaging on the hypersurface of constant time could be gauge dependent. 
 The question of which of the N-possible observers in the Universe sees a homogeneous and isotropic FLRW spacetime is still unaddressed~\cite{Clarkson:2011zq,Larena:2012vn}. It is important to note that observing isotropic distribution of matter on the past light cone of an observer does not immediately imply a homogeneous distribution of matter without a further assumption of the Copernican principle~\cite{Maartens:2011yx,Clarkson:2012bg}.
This work is about the cosmological fitting problem in the light of the current Hubble tension. 

Hubble tension refers to the discrepancies in the value of the Hubble rate, $H_0$, when late/early time observations are interpreted using the FLRW spacetime. For example, the SH0ES collaboration constrains $H_0$ to be $H_0=73.2 \pm 1.3 $ ${\rm{km/sec/Mpc}}$ from the intercept of the Hubble diagram of SNIa with the absolute luminosity calibrated using nearby Cepheid variables~\cite{Riess:2016jrr}, {detached eclipsing binary system in the Large Magellanic Cloud (LMC)~\cite{2019Natur.567..200P} and distance to the NGC 4258~\cite{Reid:2019tiq}}.  The Carnegie-Chicago Hubble Program (CCHP) constrains $H_0$ to be $H_0= 69.6\pm1.9$ $ {\rm{km/sec/Mpc}}$ using the information contained in the Hubble diagram at low redshift. Here, the absolute luminosity is calibrated using the Tip of the Red Giant Branch (TRGB) stars in the Hertzsprung-Russell diagram~\cite{Freedman:2019jwv}. The calibration of the absolute luminosity using different anchors by the two groups does not seem to be the source of the discrepancy but rather the ability to synchronise the zero-point of the distance modulus~\cite{Anand:2021sum}. The Baryon Acoustic Oscillation  (BAO)  spectroscopic survey however,  constrains a combination of the Alcock-Paczy\'{n}ski parameters~\cite{Alcock:1979mp} with respect to the BAO peak at an effective redshift different from zero. The Alcock-Paczy\'{n}ski parameters are then interpreted in terms of the flat FLRW model to make a determination of $H_0 = 68.6 \pm 1.1$ $ {\rm{km/sec/Mpc}}$~\cite{Anderson:2013oza,BOSS:2016wmc,Philcox:2020vvt}. There are a whole lot of other determinations of $H_0$ with values falling into either within BAO or SH0ES categories, for details see~\cite{DiValentino:2021izs}.

All these efforts assume that the FLRW spacetime provides the correct description of the Universe on all scales~\cite{Anderson:2013oza,Riess:2016jrr,Aghanim:2018eyx,Freedman:2019jwv,Philcox:2020vvt,DES:2021jns}, yet they return different determinations of $H_0$.  
We study this tension by averaging over the spacetime independent inhomogeneous generalisation of each of the observable-redshift relations used in the determination of $H_0$. We make use of the formalism developed by Kristian and Sachs~\cite{1966ApJ...143..379K} to derive these equations in the limit of small redshifts. We fit  FLRW model at fixed redshift to the corresponding   monopole (past-light cone average) of the generalised observable-redshift relations.
We made use of the $1+3$ covariant decomposition formalism developed in~\cite{Ellis:1966ta,Ellis:1998ct} to decompose the observable-redshift relations into covariant irreducible units. In addition, we made use of the observer past-light cone moment decomposition formalism developed in~\cite{Pirani2,Thorne1980} to extract the monopole in a coordinate independent way. This allows us to obtain the best-fit Hubble rate and the deceleration parameter, $q_0$, using each of the observable-redshift relations by matching the monopole of the arbitrary inhomogeneous model to that of the homogeneous and isotropic FLRW spacetime order by order in redshift.

We find that the difference between the Hubble rate obtained from fitting the local distance modulus for the SNIa and the Alcock-Paczy\'{n}ski parameters to the FLRW model is proportional to the square of the shear tensor associated with a geodesic observer in our local group. In a perturbed FLRW spacetime, this term is proportional to the variance in the mass density field fluctuation smoothed at scale $R$. 
We show that the smoothing scale $R$ corresponds to a scale where the geometry of the local observer decouples from the large scale expansion of the Universe(i.e the radius of the zero-velocity surface). We argue that the smoothing scale must be set to this value in order to avoid the caustics or the conjugate point at the boundary of our local group~\cite{Witten:2019qhl}. 
Taking the mass of our local group to be $M_{\rm{LG}} \sim(10^{11}-10^{12})M_{\odot}$ \cite{Li:2007eg,Carlesi:2016eud,Li:2021sqb} gives a smoothing scale of $R\sim(0.8- 1.2) {\rm{Mpc}} $. The translates to about (9-12)\% discrepancy in the Hubble rate and this is the amount needed to resolve the supernova absolute magnitude tension as well~\cite{Efstathiou:2021ocp}.
 In addition, we find that the deceleration parameter, $q_0$ obtained from cosmography is also impacted by cosmic structures.  We find a discrepancy in the determination of $q_0$. The value we find is in agreement with the recent measurement of $q_0$ from Pantheon supernova sample~\cite{Camarena:2019moy}.  It is important to note that monopole of the low redshift Taylor series expansion of the area distance differs significantly from the full expression~\cite{Umeh:2022hab}, therefore, this result may not be trusted.  We argue that it is rather an indication of a break-down of the low redshift Taylor series expansion since the determination using the full expression gives a much lower value~\cite{Riess:2021jrx}.

The rest of the paper is organised as follows: we describe the underlying philosophy behind the cosmological fitting problem in section \ref{fittingproblem}. This is followed by a description of the generalised spacetime independent inhomogeneous models for the area/luminosity distance in sub-section \ref{sec:covariantdecomp}. We describe how to fit inhomogeneous models of various observables to the FLRW model in section \ref{sec:fittingproblem}. We specialised the discussion to a perturbed FLRW model in section \ref{sec:perturbation}. We discuss the existence of a causal horizon and how to identify it in section \ref{sec:causalhorizon}. We conclude in section \ref{sec:conc}. We provide details on how the generalised inhomogeneous observable-redshift relations were derived in Appendix \ref{sec:KristainandSachs}.

{\bf{Cosmology}}: We adopt the following values for the cosmological parameters of the standard model \cite{Aghanim:2018eyx}: the dimensionless Hubble parameter, $h = 0.674$, baryon density parameter, $\Omega_{\rm b} = 0.0493$,  dark matter density parameter, $\Omega_{\rm{cdm}} = 0.264$,  matter density parameter, 
$\Omega_{\rm m} = \Omega_{\rm{cdm}} + \Omega_{\rm b}$,  spectral index, $n_{\rm s} = 0.9608$,  and the amplitude of the primordial perturbation, $A_{\rm s} = 2.198 \times 10^{-9}$.

\section{Introduction to the cosmological fitting problem}\label{fittingproblem}

There are two approaches for building a model of the Universe;  the pragmatic  and observational approaches. The pragmatic approach makes assumptions about the geometry of the Universe and then use observations to validate those assumptions.
For example, the standard model of cosmology is built on {\emph{a priori}} assumption that all physical quantities measured by a comoving observer are spatially homogeneous and isotropic (cosmological principle), This assumption restricts a set of all possible spacetimes of the Universe to the FLRW~\cite{2010fimv.book..267S}.  The current effort in cosmology is mainly directed towards making a very precise determination of the free parameters of this model~\cite{Aghanim:2018eyx,Sugiyama:2021axw,Zhao:2021ahg}. 
While the alternative observational approach uses  observational data from our past light cone such as apparent luminosities, angular diameters and number count of sources without assuming  the cosmological principle  to construct the geometry of the Universe\cite{McClure:2007hy, Hellaby:2008pp,Bolejko:2011ys}. The observational approach encounters difficulties due to the absence of the initial data~\cite{Lu:2007gr}. However,  it holds a huge  promise that when all the issues are resolved,  it will  not only provide us with  exact geometry of the Universe, it will also allow us to quantify homogeneity scale if anything like that exists.

The cosmological fitting approach we describe here is an intermediate approach between these two approaches~\cite{Ellis:1984bqf}.  It does not assume  {\emph{a priori}} that the FLRW spacetime describes the observable Universe accurately at all times and at all distances, rather it considers it as a fiducial model of the Universe.  For example, we assume that the fiducial model is specified by 
\begin{eqnarray}
\bar{U} = \left\{\bar{M}, \bar{g}_{ab} ,\bar{u}^a, \bar{\rho} , \bar{N} , \bar{\mu},\bar{a}_b, \bar{\alpha}_{\p}, \bar{\alpha}_{\bot} \right\}\,,
\end{eqnarray}
where $\bar{M}$ is Riemannian manifold, $ \bar{g}_{ab}$ is the metric, $\bar{u}^a$ is the four velocity, $ \bar{N}$ is the number count of sources, $ \bar{\mu}$ is the distance-redshift relation (distance modulus), $a_b$ is the intercept of the Hubble diagram,  $\bar{\alpha}_{\p}$  and $\bar{\alpha}_{\bot}$ are the radial  and orthogonal components of the Alcock and Paczy\'{n}ski parameters~\cite{Alcock:1979mp}. Then, it takes some hints from the observational approach such that it assumes that there exists a  model of the observed Universe which gives a realistic representation of the Universe including all inhomogeneities down to some specified length scale $|\x_1-\x_2|>R$
\begin{eqnarray}
U = \left\{M, g_{ab} ,u^a, \rho , N , \mu,a_b, \alpha_{\p}, \alpha_{\bot}\right\}\,.
\end{eqnarray}
The task then is to determine a best fit model of the Universe given $U$. Just like in the observational approach, the most optimal procedure for obtaining the best fit model  is to fit the observed data or observables  on the past light cones $C^{-}( \bar{p})$ and $C^{-}( p)$ of points $\bar{p}$, $p$ in $\bar{U}$ and $U$ respectively. %Thus we focus only on observables on the past
Since $\bar{U}$ is homogenous and isotropic, any point is equally likely, for the points in $U$, we smooth over angular dependence of any observable, $X$,  {associated with  $U$},  on the past light cone of an observer with four velocity $u^a$ within a sphere of constant redshift 
 \begin{eqnarray}
 \averageA{X(z,{\n})} %=  \int {\d}  {\Omega} \,X(z,{\n}) 
 = \frac{1}{4\pi} \int \d^2\Omega X(z,{\n}) =   \int {\d}  {\n} \,X(z,{\n})\,,
\end{eqnarray}%$\Omega$ is the area of the sky and
where  ${\n}$ is line of sight direction. We assume that the redshift is monotonic for points separated by distance greater than $R$.
%It is important that the redshifts in both the fiducial model and the realistic inhomogeneous model correspond to the coherent  Hubble flow.
We compare the monopole of the observable $X$ obtained by smoothing out anisotropies on the past null cone $C^{-}(p)$ to the corresponding prediction by the fiducial model of the same observable.   The free parameters of the fiducial model, i.e  $\bar{U}$ are then adjusted  to obtain the best-fit  value to the angular average of the observable in the lumpy Universe. In principle, we evaluate the $\chi^2$ of any observable $X$:
\begin{equation}
\chi^2_{X} = \sum_{i} \left[ \frac{\averageA{X_{i}} - \bar{X}\left(z_{i}| H^X_{0},q^X_{0}\right)}{\sigma^{X}_{i}}\right]^2\,,
\end{equation}
where $\sigma^{X}_{i}$ is the covariance, in a single parameter, it reduces to the variance, $H_0$ and $q_0$ are the Hubble rate and the deceleration parameter respectively.
Then we find the optimum values of $H^{X}_{0}$ and $q^{X}_{0}$ by minimising the $\chi^2_{X}$  with respect to $H^{X}_0$ and $q^{X}_0$
\begin{eqnarray}
\frac{\partial \chi_{X}^2}{\partial H^{X}_0} = 0 \,, \qquad \qquad {\rm{and}} \qquad \qquad \frac{\partial \chi^2_{X}}{\partial q^{X}_0} = 0\,.
\end{eqnarray}
In order to  build a complete picture of the best-fit model, the ideal thing we will be  to obtain the optimum values of $H_{0}$ and $q_{0}$ by minimaxing the joint $\chi^2 = \chi^2_{\mu} + \chi^2_{a_{b}} + \chi^2_{\alpha_{\p}} + \chi^2_{\alpha_{\bot}} + \cdots$  with respect to $H_0$ and $q_0$ simultaneously. {We do not consider this here because our focus is to identify the source of  tension in the inferred Hubble rate.  This is best done by independently fitting the late and early Universe observables to their corresponding FLRW space models.}

This intermediate approach offers tremendous advantages over the other two approaches, such as;
\begin{itemize}

\item {\tt{Clear description of the relevant physics:}} {It makes apparent the geometrical and physical interpretation of the FLRW models we use because it provides a clear link between the highly symmetric fiducial model  and more realistic descriptions of the lumpy Universe we observe.}

\item {\tt{Delineation of scales:}} It helps to  establish the  appropriate scale of validity of the best-fit model of the Universe. That is it helps to address statement that the Universe can be regarded as almost FLRW Universe if averaged out over a specified length scale. 

\item {\tt{Guidance in the presence of tension:}}
{It does not only enable one  to determine the best-fit FLRW Universe model, it allows to predict the presence of tension in the model parameters. For example, under this framework tensions appear in the  model parameters when the fiducial model is fit to data from length scales  or time in the evolution of the Universe where the background FLRW spacetime is not applicable(i.e nonlinear scales). We discuss this point in greater detail in the subsequent sections.}
% along side  details of the fitting  procedure  between the two models with discrepant predictions for the values of the fiducial model. 

\item {\tt{Wider application:}} It is  possible to use it repeatedly; that is, to  consider  which of the  lumpy Universe models $U$ and another lumpiness  model, say  $U'$ give an even better description of the real Universe than $U$, i.e which one of them describes the inhomogeneities at an even more detail.
\end{itemize}

%Finally construct the best-fit spacetime itself inside and outside  the null cone  from  the initial data on the null cone, and, insofar  as is possible, estimate how good the fit is off the light cone

\subsection{Inhomogeneous distances in a realistic model of the  Universe}\label{sec:covariantdecomp}

We work in the geometric optics limit i.e we assume that the wavelength of photon is small compared to the radius of curvature of the Universe. In this limit,  null geodesics describes photon propagation and  a tangent vector to the null geodesic is given by $k^a={\d { x}^{a}}/{\d\lambda}$ and it satisfies
 \begin{equation}
 {k}_a{k}^a=0\,\qquad {\rm{and}}\qquad {k}^b{\nabla}_b{ k}^a=0\,.
 \end{equation}
where $\lambda$ is the affine parameter. When initialised at the observer position, it increases monotonically~\cite{Perlick:2010zh}.  $\nabla$ is the covariant derivative of the physical spacetime of the Universe. 
With respect  to a set of fundamental  comoving observers with the average  4-velocity $u^a$, we can decompose ${ k}^{a}$  into parallel and orthogonal components:  
\begin{eqnarray}\label{4vector}
{ k}^a=(-{ u}_b{ k}^b)\left({ u}^a-{ n}^a\right)={E}\left({ u}^a-{ n}^a\right),%~~~n_an^a=1,~n_au^a=0, ~{E}=-{ u}_b{ k}^b.
\end{eqnarray}
where $n^a$ is the  Line of Sight (LoS) spatial direction vector with a normalization $n_an^a=1,$ $E=-{ u}_b{ k}^b$ is the photon energy measured by $u^a$. $n^a$ is orthogonal to $u^a$; $n_au^a=0$.  Note that our choice of $n^a$ is opposite to the direction of photon propagation. $k^a$ is pointing along an {incoming} light ray on the past light cone of the observer.  
The apparent magnitude of any source, $m$,  is related to the observed flux density $F_{d_L}$ measurement at a given luminosity distance $d_{L}$  in a given spectral filter according to 
\begin{eqnarray}
m=-2.5\log _{10}\left[F_{d_{L}}\right]\,.
\end{eqnarray}
The absolute magnitude, $M$, of the same source is defined as the apparent magnitude  measured at a  distance $D_{F}$
\begin{eqnarray}
M=-2.5\log _{10}\left[{F_{D_F}}\right]\,,
\end{eqnarray}
where $F_{D_F}$ is  called the reference flux or the zero-point of the filter~\cite{Rizzi:2007ni,2015PASP..127..102M,Madore:2021ktx}. The observed flux density at both distances obey the inverse square law with the luminosity distance, $d_{L}$:
${F_{d_L}}/{ F_{D_{F}}} = \left[{D_{F}}/{d_L}\right]^2.$
The distance modulus is defined as the difference between $m$ and $M$
\begin{eqnarray}\label{eq:distancemod}
m - M  = - 2.5 \log \left[\frac{ F_{d_L}}{F_{D_{F}}}\right] = 5 \log \left[\frac{d_{L}}{D_F}\right]
= 5 \log \left[\frac{d_{L}}{[\rm{pc}]}\right] - 5 \,.
\end{eqnarray}
In the last equality, we set  $D_{F} =10 \,{\rm{pc}}$ for historical reason.  In this case, it means that the absolute  magnitude  is  the apparent  magnitude  if the telescope is placed at a distance of 10 pc. In cosmology however,  $D_{F}$ is set to  $D _{F}=1 \,{\rm{Mpc}}$  leading to 
\begin{eqnarray}
{\mu}(z,{\n})&=& m(z,{\n})-M
= 5 \log \left[\frac{d_{L}}{[\rm{Mpc}]}\right] +25 \,,% 25 + 5 \log_{10} {d}_{L}(z,{\n})\,,
\label{eq:mag-redshift-relation}
\end{eqnarray}
where ${d}_L$ is in  the  units of ${\rm{Mpc}}$.  This  implies that for cosmological purposes, the consistently calibrated cosmic distance ladder would have a  reference flux  determined at $1 {\rm{Mpc}}$. 
The generalised  coordinate independent expressions for the area and the luminosity distance at low redshift are 
 (see appendix \ref{sec:KristainandSachs} for details of the derivation) \cite{1966ApJ...143..379K,Heinesen:2020bej}
 \begin{eqnarray}\label{eq:GendA}
d_A(z,{\n})&=&\frac{z}{[K^cK^d\nabla_cu_d]_o}-\frac{1}{2}\left[\frac{K^cK^dK^e\nabla_e\nabla_du_c}{\left(K^cK^d\nabla_cu_d\right)^3}\right]_oz^2+\mathcal{O}(z)^3\,,
\\
d_L(z,{\n})&=&\frac{z}{[K^cK^d\nabla_du_c]_o}\left\{1+\frac{1}{2}\left[4-\frac{K^cK^dK^e\nabla_e\nabla_du_c}{\left(K^cK^d\nabla_cu_d\right)^2}\right]_oz +  \mathcal{O}(z^2) \right\},\label{eq:Glumdist}
\end{eqnarray}
where $K^a$ is a normalized  null vector: $K^a=u^a-n^a.$ The associated null geodesic tangent vector is given by $k^a=(1+z)K^a$.  
 In general, $d_{L}$ is related to  $d_A $ via distance duality relation: $d_{L}= (1+z)^2 d_{A}$~\cite{Etheringto:1933,Ellis1971grc..conf..104E,Ellis2009}.  
 %We neglect the impact of the heliocentric peculiar motions, this will be studied elsewhere. 
 We neglect the effect of the heliocentric peculiar velocity. Its contribution will not substantially change the the determination of the Hubble rate, expecially via the local distance ladder~\cite{Peterson:2021hel}.
 The redshift in equations \eqref{eq:GendA} and \eqref{eq:Glumdist} corresponds to the cosmological redshifts.

% \subsection{Homogenous distances in a fiducial model of the Universe}
%Our plan is to obtain a best-fit FLRW model by comparing a set of FLRW distance measures to a set of generalised distance measures of a lumpy Universe.  
%To do this, we assume that there exists a set of fundamental observers with  4-velocity $u^a$ who has access to information about the lumpy Universe on his/her past lightcone. 
%We then require the observations of the lumpy Universe be  interpreted  in terms of the FLRW spacetime. 
%The key distance measures we are interested in are the area distance, $d_A$ and magnitude-redshift relation $\mu$.  Every other observable such as  the intercept of the Hubble diagram, radial  and orthogonal components of the Alcock and Paczy\'{n}ski parameters can be obtained from them. 

On the FLRW background space, equation \eqref{eq:GendA} and  \eqref{eq:Glumdist} reduce to 
\begin{eqnarray}\label{eq:FLRWdA}
\bar{d}_A(z) &=& \frac{z}{H_0}\left[1-\frac{1}{2}\left(3+q_0\right)z+\mathcal{O}(z)^2\right]\,,
\\
\label{eq:FLRWlumdist}
\bar{d}_L(z) &=&  {\; z\over H_0}
\left[ 1 + {1\over2}\left[1-q_0\right] {z} + \mathcal{O}(z)^2 \right]\,.
\end{eqnarray}
The magnitude-redshift relation is given by equation \eqref{eq:mag-redshift-relation} with the luminosity distance given by equation \eqref{eq:FLRWlumdist}.
We have truncated the Taylor series expansion at second order in redshift expansion because our focus is on how the cosmic structures impact the measurement of the Hubble rate, $H_0$ and the deceleration parameters, $q_0$. It is certainly debatable whether these series expansions on arbitrary spacetime have much relation with the magnitude-redshift relation in the observable Universe. In particular, is it analytic, especially in the region where shell crossing occurs? 
These doubts are valid, but, this is how we address them:

\begin{itemize}
\item {\tt{Differentiability at $z=0$}:} We showed in \cite{Umeh:2022hab} that $d_A$ expanded up to second order in standard cosmological perturbation theory on an FLRW background spacetime is differentiable at $z =0$.  The linearly perturbed FLRW equations can be re-written in terms of the equations of the exact inhomogeneous Szekeres models~\cite{Sussman:2017otk}.  This shows that the generalised inhomogeneous  model  is also differentiable at $z=0$.
% provided that the conjugate points outside the observer/source conjugate pair are isolated.

\item {\tt{Shell crossing singularity}:} It is likely that in certain directions, the Taylor series expansion of observables will be multivalued as a light beam passes through collapsing regions, this is a valid concern. We show in section  \ref{sec:causalhorizon}, that collapsing region around the observer can be  isolated from the expanding  spacetime by setting the smoothing scale at the zero-velocity surface~\cite{Umeh:2022hab}. Similar restriction was deployed recently in an attempt to estimate equation \eqref{eq:Glumdist} from a general relativistic N-body simulation~\cite{Macpherson:2021gbh}.

\item {\tt{Convergences of the series expansion}:} Convergence of series expansion is a problem in general, in our case, it is well-known that for $z < 0.1$, equations \eqref{eq:FLRWdA} and \eqref{eq:FLRWlumdist} converge at second order in redshift expansion. It is likely that in the presence of structures, higher-order redshift correction will be needed to achieve convergence at the same redshift. We find some hints of this in~\cite{Umeh:2022hab} but a more detailed study is needed since next to leading order terms were neglected in the analysis. Currently, we are essentially interested in what happens in the limit where $z\ll 1$ and we have no reason to believe that second-order series expansion in redshift will not be enough to determine the Hubble rate. 

\end{itemize}

%\eqref{eq:GendA}\eqref{eq:lumdist} -\eqref{eq:Magnzrelation

\subsection{Covariant and multipole moment decomposition of observables}\label{eq:observables}

We are primarily interested in the multipole moment decomposition of the observable-redshift relations so that the monopole or the all-sky average can be compared to the corresponding expression based on the background FLRW spacetime. To accomplish this, we decompose $K^aK^b\nabla_au_b$ and $K^aK^bK^c\nabla_a\nabla_bu_c$ into irreducible physical quantities that live on the hypersurface orthogonal to $u^a$ using in 1+3 covariant decomposition formalism~\cite{Ellis:1966ta,EGS1968,Ellis1971grc..conf..104E,Ehlers:1993gf,Ellis:1998ct,Ellis:1990gi}.
 The irreducible decomposition of the spacetime covariant derivative of $u^a$ for geodesic observers is given by~\cite{Ellis:1998ct,Ellis:1990gi}
\begin{eqnarray}
\nabla_{b}u_{a}=\frac{1}{3}\Theta h_{ab }
+\sigma_{ab}\,, \label{eq:covdu}
\end{eqnarray}
 where $\Theta$ denotes the expansion ($\Theta >0$)/contraction $(\Theta<0)$ of the nearby geodesics associated with $u^a$, $\sigma_{ab}$ is the shear tensor, it describes the rate of change of the deformation of spacetime in the neighbourhood of the observer, $h_{ab}=g_{ab}+u_a u_b$ is the metric on the hypersurface orthogonal to the time-like $u^a$ ($u^{a}u_{a} = -1$) and $g_{ab}$ is the physical spacetime metric. The vorticity vanishes for irrotational fluids or geodesic observer.  
There is a $1$-to-$1$ mapping between all symmetric trace-free tensors of rank $\ell$ and the spherical harmonics of order $\ell$ \cite{Pirani2,1981MNRAS.194..439T,2000AnPhy.282..285G,Ellis1983455}.  
We can see this by decomposing the LoS direction vector $n^a$ at the observer position on an orthonormal tetrad basis~\cite{2000AnPhy.282..285G,Gebbie:1998fe}
\begin{equation}
n^a(\theta, \phi)=\left(0,\sin \theta\sin\phi,\sin\theta\cos\phi, \cos\theta\right)\,.
\end{equation}
In the standard spherical harmonics decomposition formalism, any function $f$ on the sky may be expanded in spherical harmonics, $Y^{\ell m}({\n})$ according to 
\begin{eqnarray}\label{eq:PSTFexpansion}% \label{eq:sph}
f({\n}) = \sum_{\ell=0}^{L} \sum_{m=-\ell}^{\ell} F^{\ell m}
	Y^{\ell m}({\n}) = \sum_{\ell=0}^{\infty} F_{A_{\ell}} n^{A_{\ell}} = F+ F_an^a + F_{ab}n^an^b + F_{abc}n^an^bn^c\,.
\end{eqnarray}
where $F^{\ell m}$ is the spherical harmonic coefficients. 
$A_{\ell} = a_1 a_2 ... a_{\ell}$ is a compound index notation that denotes number of indices, $F_{A_{\ell}}$ is the moments, it is  symmetric, trace-free and orthogonal to $u^a$, $
F_{A_{\ell}}=F_{(A_{\ell})}\,,$ $~F_{A_{\ell} ab} h^{ab}
=0\,,$ and $F_{A_{\ell} a}  u^{a}  = 0\,$ respectively.
%\end{eqnarray}
Here, $``(\cdots)"$ brackets in the subscript denote the symmetrisation overall indices and the angle brackets ``$\<\cdots\>$'' denotes PSTF part. 
In the second equality, we show an equivalent way of performing the spherical harmonic expansion in Projected Symmetric Trace-Free (PSTF) tensors in $n^{a}$basis~\cite{Pirani2,1981MNRAS.194..439T,2000AnPhy.282..285G,Ellis1983455}.
  The PSTF part of any index-tensor is given  by 
\cite{Pirani2,Spencer1970,Thorne:1980ru,Liotta2000L,Jaric2003},
\begin{eqnarray}
 {F_{\<A_\ell\> }}
=\sum_{n=0}^{[\ell
/2]}\mathcal{B}^{\ell n}h_{(a_1a_2}....h_{a_{2n-1}a_{2n}}F_{a_{2n+1}...a_\ell )}\,, \quad{\rm{with}}\quad \mathcal{B}_{\ell n}={\frac{{(-1)^n\ell !(2\ell -2n-1)!!}}{{(\ell -2n)!(2\ell
-1)!!(2n)!!}}}\,.
 \label{eq:PSTFdef} 
\end{eqnarray}
 $[\ell /2]$ means the largest integer part less than or equal to
$\ell /2$, $\ell !
=\ell (\ell -1)(\ell -2)(\ell -3)...(1)$, and $\ell !! =\ell (\ell
-2)(\ell -4)(\ell -6)...(2~~or~~1)$.
The moment and covariant decomposition of $K^aK^b\nabla_au_b$ and $K^aK^bK^c\nabla_a\nabla_bu_c$ within General Relativity for dust is given by \cite{Clarkson:2011zq,Clarkson:2011uk,Umeh:2013UCT}
\begin{eqnarray}
K^aK^b\nabla_a u_b&  =& \frac{1}{3}\Theta + \sigma_{ab}n^an^b\,,\label{eq:decompstuff}
\\
\label{eq:decomposestuff2}
K^aK^bK^c\nabla_a\nabla_b u_c  &=& \mathcal{O} +  \mathcal{O}_{a} n^a +  \mathcal{O}_{ab}n^a n^b +  \mathcal{O}_{abc}n^a n^b n^c
\end{eqnarray}
where $ \mathcal{O}$, $ \mathcal{O}_{a} $, $ \mathcal{O}_{ab}$ and $\mathcal{O}_{abc}$ are the monopole, dipole, quadrupole and octupole moments of $K^aK^bK^c\nabla_a\nabla_b u_c$
\begin{eqnarray}
 \mathcal{O}&=&\frac{1}{6}\rho + \frac{1}{3}\Theta^2 -\frac{1}{3}\Lambda 
+ \sigma_{ab}\sigma^{ab}\,,
\\
 \mathcal{O}_{a}&=&\frac{1}{3}\D_a\Theta+\frac{2}{5}\D_b{\sigma}_a^b \,,
 \\
 \mathcal{O}_{ab}&=&E_{ab}+ 2\Theta\sigma_{ab}+ 3 {\sigma^c}_{a}\sigma_{bc}\,,
\\
\mathcal{O}_{abc}&=& \D_{a}\sigma_{bc}\,.
\end{eqnarray}
Here $\rho$ is the matter density, $\Lambda$ is the cosmological constant, $E_{ab}$ is the Electric part of the Weyl tensor and {$\D_a$ is the spatial derivative on the hypersurface.} $E_{ab}$  describes how nearby geodesics tear apart from the equation. It is crucial in the discussion that follows. 
An extension to modified gravity theories and general matter fields should be straightforward.  

\section{The cosmological fitting problem}\label{sec:fittingproblem}

In this section, we consider how surveys that measure the imprints of the Baryon Acoustic Oscillation (BAO) in galaxy distribution and the apparent magnitude of the type Ia supernova are compared to a cosmological model through the Alcock-Paczy\'{n}ski parameters, the distance-redshift relation. Our key focus will be to estimate how inhomogeneity affects the determination of the Hubble rate and the deceleration parameter.

\subsection{Cosmological fitting problem: Alcock-Paczy\'{n}ski parameters}
%aused by acoustic density waves in the primordial plasma of the early Universe

The BAO scale constitutes an important link between the physics, before and around the drag and present epoch. %by its property as a statistical “standard ruler”
Similar to the Cosmic Microwave Background radiation, CMB, the plasma physics of acoustic density waves before decoupling imprints a characteristic scale in the matter distribution. The characteristic scale, i.e BAO scale appears as an excess in the probability of finding galaxies separated by some distance in the distribution of sources seen today on the sky. Since a fiducial model for the separation is required for calculating the separation between galaxies, Alcock and Paczy\'{n}ski \cite{Alcock:1979mp} introduced a clever way to recover the true separation.  For a spherical symmetric distribution of sources, the two-point correlation function, $\xi_{g}$, is given by~\cite{Nadathur:2019mct}
\begin{eqnarray}\label{eq:rescaled2pcf}
\xi_{g}(r_{12\p}, r_{12\bot})  = \frac{1}{\alpha_{\p} \alpha_{\bot}^2} \xi_{g}(\alpha_{\p}r^{\rm{fid}}_{12\p}, \alpha_{\bot} {\r}^{\rm{fid}}_{12\bot}) \,,
\end{eqnarray}
where $r_{12}$ and  $r^{\rm{fid}}_{12}$ are the true and fiducial separation between two galaxies respectively. Within the cosmological standard model treatment, the difference between the fiducial model and true model are parameterised by
\begin{eqnarray}\label{eq:BAOparams}
\bar{\alpha}_{\p} = \frac{ \bar{H}^{\rm{fid}}}{\bar{H}} \,, \qquad\qquad \bar{\alpha}_{\bot }  = \frac{\bar{d}_{A}}{\bar{d}^{\rm{fid}}_{A}}\,.
\end{eqnarray}
Here $\bar{H}^{\rm{fid}}$ and $\bar{d}^{\rm{fid}}_{A}$ are the fiducial Hubble rate and area distance respectively. $\bar{H}$ and
 $ \bar{d}_{A}$ are then adjusted to obtain the best-fit to the observed data. 
BAO is sensitive to the Hubble rate through the distortions ($\alpha_{\p}$ and $\alpha_{\bot}$)  at the survey mean redshift $z$.  The determination of the value of the Hubble rate today is made assuming a model.  The  measurement of the monopole  of equation \eqref{eq:rescaled2pcf} constrains $\alpha = \alpha_{\bot}^{{2}/{3}}\alpha{\p}^{{1}/{3}}$~\cite{Anderson:2013oza,Beutler:2014yhv}, while the quadrupole moment is most sensitive  to the  Alcock-Paczy\'{n}ski ratio $\epsilon = \alpha_{\bot}/\alpha_{\p}$~\cite{Nadathur:2019mct}.  Using the void-galaxy cross-correlation  from the BOSS survey at the mean redshift of $z =0.57$, ~\cite{Nadathur:2019mct} finds a strong constraint on $\alpha_{\bot}/\alpha{\p} =  1.016\pm0.011$ {using the CMASS galaxy sample of the BOSS DR12 data release}. This constraint when interpreted in within the $\Lambda$CDM model{(i.e  equation \eqref{eq:BAOparams} with the  sound horizon, $r_s$, set to the Planck value)} is consistent  with the Planck collaboration's  determination of $H_0$ from the analysis of the anisotropies in the cosmic microwave background (CMB)~\cite{Aghanim:2018eyx}:
\begin{eqnarray}
 H_0 = 67.4 \pm 0.5\,\, {\rm{km/sec/Mpc}}\,.
\end{eqnarray}

In a lumpy Universe, the structure of equation \eqref{eq:rescaled2pcf} will change slightly, however, it holds for spherical symmetric distribution of sources. We will discuss modifications away from this limit elsewhere. For an almost FLRW spacetime for which we are interested here, equation \eqref{eq:BAOparams} generalises as follows:
\begin{eqnarray}\label{eq:dEdz}
\alpha_{ \p} &=& \frac{\partial r_{\p}}{\partial r^{\rm{fid}}_{\p }} =  \frac{\partial r_{\p}}{\partial z}\frac{\partial z}{\partial r^{\rm{fid}}_{\p}} \simeq \frac{\d \lambda}{\d  z}\frac{\d  z}{\d \lambda^{\rm{fid} }} \,,
\\
&=&\frac{ {\bar{H}} ^{\rm{fid} }}{{H}}  \left[1
+ \frac{\sigma_{ab}}{{H}} {n}^{a}n^{b}\right]^{-1}\,,
\\
&=&\frac{ \bar{H} ^{\rm{fid} }}{{H}} \bigg[1- \frac{\sigma_{ab}}{{H}}n^a n^b +  \frac{\sigma_{ab}}{{H}} \frac{\sigma_{cd}}{{H}} n^a n^b n^c n^d  +\mathcal{O}(\sigma^4)\bigg]\,,
\end{eqnarray}
where, $H =\Theta/3$, in the last equality, This is the Hubble rate associated with the expansion of the nearby congruence in a lumpy Universe. Moreover, we assumed that the real Universe is almost statistical homogeneous and isotropic, i.e  $ {\sigma_{ab}}n^a n^b/{{H}} \ll1$. To move from the first line of equation \eqref{eq:dEdz} to the second, we made use of the propagation equation for the redshift~\cite{Umeh:2014ana}
\begin{eqnarray}\label{Redshiftprorpa1}
\frac{\d  z}{\d \lambda}%=-(1+ z)^2\left[\frac{1}{3} \Theta + \sigma_{ab}n^{a}n^{b}\right] 
&=&-(1+ z)^2\left[{H}
+ \sigma_{ab}n^{a}n^{b}\right] \,,
\qquad \qquad \frac{\d  z}{\d \lambda^{\rm{fid} }} =-(1+ z)^2 \bar{H} ^{\rm{fid} }\,.
\end{eqnarray}
Note that on the fiducial FLRW spacetime, the shear vanishes  $\sigma_{ab} = 0$.  At low redshift, the generalised counter-part to $\bar{\alpha}_{\bot}$ involves replacing the background area distance with equation \eqref{eq:GendA}.

To obtain the best-fit model from the radial component of the separation in the limit $z\to0$, we compare $\bar{\alpha}_{\p} $ to all-sky average of $\alpha_{ \p} $
\begin{eqnarray}
\bar{\alpha}{\p}  = \averageA{\alpha_{ \p} }&=&\frac{ \bar{H} ^{\rm{fid} }}{{H}} \bigg[1+  \averageA{\frac{\sigma_{ab}}{{H}} \frac{\sigma_{cd}}{{H}} n^a n^b n^c n^d  }+\mathcal{O}(\sigma^4)\bigg]\,,\label{eq:fidHubble0}
\\
&=&\frac{ \bar{H} ^{\rm{fid} }}{{H}} \bigg[1+ \frac{2}{15}\frac{\sigma_{ab}\sigma^{ab}}{H^2}+\mathcal{O}(\sigma^4)\bigg]\,.
\label{eq:fidHubble}
\end{eqnarray}
To perform the all-sky average we
decompose $\sigma_{(ab}\sigma_{cd)}$ into irreducible unit and then extract the monopole using
\begin{eqnarray}\label{eq:sigmasqpstf}
n^an^bn^cn^d\sigma_{(ab}\sigma_{cd)}=n^an^bn^cn^d\sigma_{\<ab}\sigma_{cd\>}+\frac{4}{7}n^an^b\sigma^c_{\<a}\sigma_{b\>c}+\frac{2}{15}\sigma_{ab}\sigma^{ab}.
\end{eqnarray}
This can also be done by performing the angular integration in equation \eqref{eq:fidHubble0} right away~\cite{Thorne1980}.
Comparing equation \eqref{eq:BAOparams} to equation \eqref{eq:fidHubble}, we obtain the effective Hubble rate 
\begin{eqnarray}\label{eq:BAOalpha}
H^{\alpha{\p}} = {H} \bigg[1- \frac{2}{15}\frac{\sigma_{ab}\sigma^{ab}}{H^2}+\mathcal{O}(\sigma^4)\bigg]\,.
\end{eqnarray}
The Hubble rate inferred from the constraint on $\alpha_{\p}$ is biased by $ -{2}\sigma_{ab}\sigma^{ab}/{15H^2}$. For the orthogonal component, $\alpha_{\bot}$, we focus on $d_A$, since the fiducial model is the same, it reduces to comparing equation \eqref{eq:FLRWdA} to the all-sky average of equation \eqref{eq:GendA} at a fixed redshift. At the leading order in redshift, the effective Hubble rate from fitting the monopole of area distance in a lumpy Universe to the FLRW spacetime is given by
\begin{eqnarray}\label{eq:H0_lum}
\frac{1}{H_0^{{d}_A}}&=&\averageA{\frac{1}{{H}_{0}}\left(1+ \frac{\sigma_{ab}}{{H}_{0}}n^an^b\right)^{-1}\bigg|_{z=z_o} }
=\frac{1}{{H}_0}\bigg[1- \frac{\sigma_{\<ab\>}}{{H}_0}n^a n^b +  \frac{\sigma_{(ab}}{{H}_0} \frac{\sigma_{cd)}}{{H}_0} n^a n^b n^c n^d  +\mathcal{O}(\sigma^4)\bigg]\,,
\end{eqnarray}
where $  \Theta_o=3{H}_{0} $ is the monopole component of $K^aK^b\nabla_au_b$.   Again,  we expanded $\left(K^aK^b\nabla_au_b\right)^{-1}$ up to second order in $ \sigma_{ab}n^a n^b/H_{0} \ll 1$.
Using equation \eqref{eq:sigmasqpstf} again and noting that  $\sigma^{a}{}_{a} =0$, we find
 \begin{eqnarray}\label{eq:H0dl}
\frac{1}{H_0^{d_A}}\equiv\frac{1}{{H}_{0}}\left[1 +\frac{2}{15}\frac{\sigma_{ab}\sigma^{ab}}{{H}^2}\bigg|_{z_o}+\mathcal{O}(\sigma^4)\right]\,.
\end{eqnarray}
The Hubble rate inferred from the area distance by fitting the area distance information from the lumpy Universe to an FLRW background spacetime is given by
\begin{eqnarray}\label{eq:HOdA}
{H_0^{d_A}}
= {H}_{0}\left[1 -\frac{2}{15}\frac{\sigma_{ab}\sigma^{ab}}{{H}^2}\bigg|_{z_o}+\mathcal{O}(\tilde\sigma^4)\right]\,.
\end{eqnarray}
Furthermore, the evolution of $\sigma_{ab}$ is sourced by the electric part of the Weyl tensor, ${E}_{ab}$~\cite{Ellis:1990gi,Ellis:1998ct}
\begin{eqnarray}
\frac{{\rm{D}}{\sigma}_{ab}}{\rm{D}\tau}&=& - \frac{2}{3} \Theta {\sigma}_{ab} - {\sigma}^{c}{}_{\<a}{\sigma}_{b\>c}  - {E}_{ab}\,.
\end{eqnarray}
 Again, ${E}_{ab}$ contains information about the tidal forces due to gravity, it represents how nearby geodesics tear apart from each other~\cite{Ellis:1990gi}. In an almost FLRW spacetime (i.e small perturbation on top of an FLRW spacetime), ${\sigma}_{ij}$ is related to $E_{ij}$ in comoving spacetime 
\begin{eqnarray}\label{eq:sigma_Epart}
{\sigma}_{ij}(\eta, {\x}) &\sim&  -  \int_{\eta_{\rm{ini}}}^{\eta}  {d\eta'} {E}_{ij}(\eta', {\x}') \,.
\end{eqnarray}
In this form, it is straightforward to see that $\sigma_{ij}$ describes the cumulative effect of the local geometry deformation from the initial seed time to the time of observation. It also helps to understand why $\sigma_{ij}$ characterises the cosmic web more transparently than $E_{ij}$~\cite{2012MNRAS.425.2049H,Forero-Romero:2014fna}. %~\cite{Hahn:2006mk,Libeskind:2017tun}

For the deceleration parameter today,   we compare equation \eqref{eq:FLRWdA} to the monopole of equation  \eqref{eq:GendA} at the second order in redshift 
\begin{eqnarray}\label{eq:defdAq0}
\frac{\left(3+q^{d_A}_0\right)}{H^{d_A}_0} &=&\averageA{\frac{K^aK^bK^c\nabla_a\nabla_bu_c}{(K^aK^b\nabla_au_b)^3}}\,.
\end{eqnarray}
Decomposing the denominator in terms of the expansion and shear tensor and expressing the result into irreducible unit, we found
\begin{eqnarray}
{\left(K^aK^b\nabla_au_b|_0\right)^3 } &=&\frac{1}{27}\Theta^3+\frac{1}{3}\Theta^2\sigma_{ab}
+ \sigma_{ab}\sigma_{cd} \Theta n^an^b n^c n^d  + \mathcal{O}(\sigma^3)\,,
\\
 &=&\Hcub  + \Hcub_{ab} n^an^b + \Hcub_{abcd}n^an^bn^cn^d + \mathcal{O}(\sigma^3)\,.
 \label{eq:Hcubed}
\end{eqnarray}
In the second equality, we have introduced the first few multipole moments as
\begin{eqnarray}
\Hcub &=& H^3 +\frac{2}{15} {\Theta  \sigma_{ab}\sigma^{ab}} = H^3\left(1 + \frac{6}{15}\frac{\sigma_{ab}\sigma^{ab}}{H^2}\right) \,,
\\
\Hcub_{ab} &=& \frac{1}{3}{\Theta^2\sigma_{ab}} +  \frac{4}{7}{\sigma_{\<a}^c\sigma_{b\>c}\Theta}\,,
 \\
\Hcub_{abcd}&=& {\sigma_{\<ab}\sigma_{cd\>}\Theta}\,.
\end{eqnarray}
Here $\Hcub $ is the monopole, $\Hcub_{ab}$, and $ \Hcub_{abcd}$ are quadrupole and the Hexadecapole moments of $\left(K^aK^b\nabla_au_b|_0\right)^3 $ respectively. 
Using equation \eqref{eq:decomposestuff2} and \eqref{eq:Hcubed} we find the best-fit deceleration parameter is given by
\begin{eqnarray}
q^{d_A}_0 &=& -3 + \frac{H^{d_A}_0}{\Hcub}\bigg[
\mathcal{O}\left(1 +  \frac{2}{15} \frac{ \Hcub_{ab}\Hcub^{ab}}{\Hcub^2}\right) - \frac{2}{15} {\frac{ \Hcub_{ab}}{\Hcub}\mathcal{O} ^{ab}} \bigg]\bigg|_{z=z_0}\,,
\\
&\approx&\frac{1}{H^2}\bigg\{\frac{1}{6}\rho  -\frac{1}{3}\Lambda 
+\frac{3}{5}\sigma_{ab}\sigma^{ab}
+\frac{2}{3}  \bigg(\frac{1}{6}\rho  -\frac{1}{3}\Lambda 
\bigg)\frac{\sigma^{ab}\sigma_{ab}}{H^2}
%\\ &&
-\frac{2}{5}  \frac{\sigma^{ab}E_{ab}}{H}
\bigg\}\bigg|_{z=z_0}\,.
\label{eq:qodA}
\end{eqnarray}
Given that the distance duality relation or Etherington reciprocity theorem holds, the Hubble and deceleration parameters obtained from the monopole of luminosity distance must correspond to the Hubble rate given in equations \eqref{eq:HOdA} and \eqref{eq:qodA} respectively~\cite{MacEllis70}. We checked and find that this holds, thus for the Hubble rate ${H_0^{d_L}} = {H_0^{d_A}} $ and the deceleration parameter $q^{d_L}_0 =q^{d_A}_0$.

\subsection{Cosmological fitting problem: Distance modulus}\label{sec:distancemod}

The Carnegie-Chicago Hubble Program, CCHP, uses the distance modulus with the luminosity distance given by the flat FLRW spacetime (equation \eqref{eq:FLRWlumdist}) to estimate $H_0$ and $q_0$ \cite{Freedman:2019jwv}. 
\begin{eqnarray}\label{eq:mudL}
\mu(z,H_0,q_0) &=& m - M\,,
\\
&=& 5\log_{10} d_{L} + 25 =5\log_{10} \left[\frac{z}{H_0}\left(1 + \frac{1}{2}(1-q_0) z + \mathcal{O}(z^2)\right)\right] + 25\,.
\label{eq:mudL1}
\end{eqnarray}
One unique feature of the CCHP approach is that it uses the Tip of the Red Giant Branch stars, TRGB, to calibrate the SNIa samples. 
According to \cite{Freedman:2021ahq}, the best-fit $H_0$ depends crucially on the accuracy with which the absolute magnitude of the SNIa is calibrated
\begin{eqnarray}\label{eq:absMag}
M = m - \mu_{0 }^{\rm{TRGB}}\,,
\end{eqnarray}
where $m$ is the apparent magnitude of the peak of the SNIa light curve for a given subsample {of SNIa co-located with the TRGB}, $\mu_{0 }^{\rm{TRGB}}$ is the true calibrator distance modulus. After calibrating $M$~\cite{Freedman:2019jwv,Freedman:2020dne},  equation \eqref{eq:mudL} is used for the SNIa sample in the Hubble flow (i.e within the redshift range ($0.03\le z\le 0 .4$)). 
CCHP usually put uninformative prior({the value of $q_{0} $ does not vary in the model  estimates of distance to each of the SNIa in the sample}) on $q_{0} = -0.53$ to determine the Hubble rate to be~\cite{Freedman:2019jwv} 
\begin{eqnarray}
H_0 = 69.6\pm1.9\,\, {\rm{km/sec/Mpc}}\,.
\end{eqnarray}

%CCHP  SNe Ia sample 

The generalised form of the distance modulus is obtained by replacing $d_{L}$ given in  \eqref{eq:FLRWlumdist} with the generalised form given equation \eqref{eq:Glumdist}.  Then compare the all-sky average to equation \eqref{eq:mudL}
\begin{eqnarray}
\averageA{\mu} = \averageA{m}-\averageA{M}  &=& {5} \averageA{ \,\log_{10}\left[{d_L(z,{\n}}\right]}+25
 \label{eq:mu_averaging}
\end{eqnarray}
 Comparing equation \eqref{eq:mu_averaging} to equation \eqref{eq:mudL} requires that we perform the angular integration over  logarithm of $d_L$ %i.e $  \log_{10} [d_L(z,{\n})]$. 
 \begin{eqnarray}
\averageA{\log_{10}d_L(z,{\n})}&=&- \averageA{\log_{10}[K^cK^d\nabla_du_c]_o} + \averageA{ \log_{10}\left[  \hat{d}_{L}(z,{\n}) \right]}\,,
\end{eqnarray}
 where we have used the quotient rule for logarithm to re-write equation \eqref{eq:Glumdist}.
We have also  introduced the generalised form of the normalised luminosity distance
  \begin{eqnarray}
 \hat{d}_{L}(z,{\n}) &=& z\left\{1+\frac{1}{2}\left[4-\frac{K^cK^bK^a\nabla_a\nabla_bu_c}{\left(K^cK^b\nabla_cu_b\right)^2}\right]_0z   + \mathcal{O}(z^2)\right\}\,.
 \label{eq:mag-red-relation3}
\end{eqnarray}
 In order to simplify $\averageA{\log_{10}[K^cK^d\nabla_du_c]_o} $ further,  we use the irreducible decomposition of $K^aK^b\nabla_a u_b$ 
\begin{eqnarray}\label{eq:KKDu-decomp}
\log_{10} \left[K^aK^b\nabla_a u_b\big|_{0} \right]&=& \log_{10}\left[H_0 \left( 1 + \frac{\sigma_{ab}}{H_0}n^an^b\right) \right] 
= \log_{10} H_0 + \log_{10} \left( 1 + \frac{\sigma_{ab}}{H_0}n^an^b\right) \,.
\end{eqnarray}
For $\sigma_{ab}n^an^b/H_0 \ll1$, we can expand the second term in Taylor series  and use equation \eqref{eq:sigmasqpstf} to obtain the PSTF part
\begin{eqnarray}
\log_{10} \left( 1 + \frac{\sigma_{ab}}{H_0}n^an^b\right)  & =& \frac{1}{\log10} \left[  \frac{\sigma_{ab}}{H_0}n^an^b -\frac{1}{2}
 \frac{\sigma_{ab}}{H_0} \frac{\sigma_{cd}}{H_0}n^an^b n^cn^d + \mathcal{O} \left(\sigma\right)^3\right]\,,
 \\
 & =& \frac{1}{\log10} \bigg[  -  \frac{1}{15}\frac{\sigma_{ab}\sigma^{ab}}{H_0^2}+  \left(\frac{\sigma_{ab}}{H_0} - \frac{2}{7}\frac{\sigma^c_{\<a}\sigma_{b\>c}}{H_0^2}\right) n^{\<a}n^{b\>}
 %\\ \nonumber &&
 -\frac{1}{2}\frac{\sigma_{\<ab}\sigma_{cd\>}}{H^2_0}n^{\<a}n^bn^cn^{d\>}
 + \mathcal{O} \left(\sigma\right)^3\bigg]\,.
\end{eqnarray}
Now it {straight forward} to perform the all-sky average to obtain the monopole  
 \begin{eqnarray}\label{eq:AverageKKDu}
 \averageA{\log_{10} \left[K^aK^b\nabla_a u_b\big|_{0}\right]} & =&  %\log_{10} H_0-  \frac{1}{15} \frac{1}{\log_{e}10}\frac{\sigma_{ab}\sigma^{ab}}{H_0^2}= 
  \log_{10} H_0-  \log_{10}\left[1+\frac{1}{15}\frac{\sigma_{ab}\sigma^{ab}}{H_0^2} \right] \,.
 \end{eqnarray}
Extracting  the monopole of $\averageA{ \log_{10}\left[  \hat{d}_{L}(z,{\n}) \right]}$ is little more algebraically more involved. We go through it step-by-step. 
Firstly, we  decompose $\hat{d}_L(z,{\n})$ in multipole moments
\begin{eqnarray}
\hat{d}_L(z,{\n})&=&\sum_{\ell=0}^\infty \hat{d}_{A_\ell}(z) n^{\<A_\ell\>}
= \hat{d}^{L}_{0} + \hat{d}^L_an^a + \hat{d}^{L}_{ab}n^an^b + \hat{d}^{L}_{abc}n^an^bn^c + \mathcal{O}\left(  \hat{d}^{L}_{A_{\ell>3}}\right)\,,
\end{eqnarray}
where $\hat{d}^{L}_{0} $, $ \hat{d}^L_a$ and $\hat{d}^{L}_{ab}$ are the monopole, dipole and quadrupole moments of the  Hubble rate-normalised luminosity distance respectively. Then by factoring out the monopole of $\hat{d}_{L}$ and requiring that $ \hat{d}^{L}_{A_{\ell>1}}/\hat{d}^{L}_{0} \ll1$  allows to  expand the anisotropic part on the background of its monopole moment
\begin{eqnarray}
  \log_{10} (\hat{d}_L(z,{\n})&=&
  \log_{10} {\hat{d}^{L}_{0}}+ \frac{1}{\log10}\bigg[ \frac{\hat{d}^{L}_{a}  }{\hat{d}^{L}_{0}}n^{a}+ \frac{\hat{d}^{L}_{ab}  }{\hat{d}^{L}_{0}}n^{a}n^{b}
%\\ \nonumber && 
 -\frac{1}{ 2}\left( \frac{\hat{d}^{L}_{a}}{\hat{d}^{L}_{0}}\frac{\hat{d}^{L}_{b}}{\hat{d}^{L}_{0}}  n^an^b + \frac{\hat{d}^{L}_{ab}}{\hat{d}^{L}_{0}}\frac{\hat{d}^{L}_{cd}}{\hat{d}^{L}_{0}}  n^an^bn^{c}n^{d}   \right)+ \mathcal{O}\left(\frac{\hat{d}^L_{A_\ell}}{\hat{d}^{L}_{0}}\right)^3 
 \bigg] \,.\label{eq:multipolesoflogdL}
\end{eqnarray}
Then taking the all sky average  leads to 
\begin{eqnarray}
\averageA{  \log_{10} (\hat{d}_L(z,{\n})} &=&
  \log_{10} {\hat{d}^{L}_{0}}- \frac{1}{2} \frac{1}{\log10}\left( \frac{2}{15}   \frac{\hat{d}^{L}_{ab}}{\hat{d}^{L}_{0}}\frac{\hat{d}_{L}^{ab}}{\hat{d}^{L}_{0}}  +\frac{1}{3}
  \frac{\hat{d}^{L}_{a}}{\hat{d}^{L}_{0}}\frac{\hat{d}_{L}^{a}}{\hat{d}^{L}_{0}} \right) + \mathcal{O}\left(\frac{\hat{d}^L_{A_\ell}}{\hat{d}^{L}_{0}}\right)^3 
\,.\label{eq:monopoledL}
\end{eqnarray}
We neglect the  contribution of the second term {since their contribution is second order in redshift $\mathcal{O}(z^2)$}
\begin{equation}
 \frac{\hat{d}^{L}_{ab}}{\hat{d}^{L}_{0}}\frac{\hat{d}_{L}^{ab}}{\hat{d}^{L}_{0}} =  \frac{\hat{d}^{L}_{a}}{\hat{d}^{L}_{0}}\frac{\hat{d}_{L}^{a}}{\hat{d}^{L}_{0}}   \approx 0\,.
 \label{eq:neglected}
\end{equation}
%{From equation \eqref{eq:mag-red-relation3}, corrections to the Hubble rate-normalised luminosity distance is quadratic in redshift. }
%We explicitly evaluate these in the Appendix. 
We  can now make use of the product rule for logarithms to re-write $\averageA{\log_{10} \hat{d}_{L}(z,{\n})} $ as
\begin{eqnarray}
%\averageA{\log_{10} \hat{d}_{L}(z,{\n})} &=& \log_{10} cz + \averageA{\log_{10} \left\{1+\frac{1}{2}\left[4-\frac{K^cK^bK^a\nabla_a\nabla_bu_c}{\left(K^cK^b\nabla_cu_b\right)^2}\right]_0z   + \mathcal{O}(z^2)\right]}\\
\averageA{\log_{10} \hat{d}_{L}(z,{\n})}  &=&\log_{10} cz +\log _{10}\bigg\{1+ \frac{1}{2}\left[4- \averageA{
\frac{K^cK^bK^a\nabla_a\nabla_bu_c}{\left(K^cK^b\nabla_cu_b\right)^2}}z\right]+ \mathcal{O}(z^2)\bigg\}
\label{eq:KKKuuKKu}
\end{eqnarray}
We  simplify this further by first  decomposing the denominator  into irreducible parts
\begin{eqnarray}
\left(K^aK^b\nabla_au_b|_0\right)^2&=&  \frac{1}{9}\Theta^2 + \frac{2}{3} n^an^b \sigma_{ab} \Theta + n^an^bn^cn^d \sigma_{ab}\sigma_{cd}\,,\\
%\left(K^aK^b\nabla_au_b|_0\right)^2&=& \frac{1}{9}\Theta^2+\frac{2}{15}\sigma_{ab}\sigma^{ab}+ n^{\<a}n^{b\>}\left(\frac{4}{7}\sigma_{\<a}^c\sigma_{b\>c}
%+ \frac{2}{3}\Theta\sigma_{\<ab\>}\right)+n^{\<a}n^bn^cn^{d\>}\left(\sigma_{\<ab}\sigma_{cd\>}\right) \qquad \\&&
&=&\Hsq+n^{\<a}n^{b\>}\Hsq_{\<ab\>}+n^{\<a}n^bn^cn^{d\>}\Hsq_{\<abcd\>}\,,
\label{decomposestuff3}
\end{eqnarray}
where $\Hsq$ is the monopole, $\Hsq_{\<ab\>}$ is the normalised quadrupole and $\Hsq_{\<abcd\>}$ is the hexadecapole
\begin{eqnarray}
\Hsq&=& \frac{1}{9}\Theta^2+\frac{2}{15}\sigma_{ab}\sigma^{ab} = H^2\left( 1 + \frac{2}{15}\frac{\sigma_{ab}\sigma^{ab}}{H^2} \right)\,,
\\
\Hsq_{\<ab\>}&=&\frac{4}{7}\sigma_{\<a}^c\sigma_{b\>c}
+ \frac{2}{3}\Theta\sigma_{\<ab\>}\,,
\\
\Hsq_{\<abcd\>}&=&\sigma_{\<ab}\sigma_{cd\>}\,.
\end{eqnarray}
Therefore, the monopole of the argument of the second term in equation \eqref{eq:KKKuuKKu} becomes
\begin{eqnarray}
  \averageA{\frac{K^cK^bK^a\nabla_a\nabla_bu_c}{\left(K^cK^b\nabla_cu_b\right)^2}\bigg|_0}&=&\frac{1}{\Hsq}\bigg[
\mathcal{O}\left(1 +  \frac{2}{15}  \frac{ \Hsq_{ab}\Hsq^{ab}}{\Hsq^2}\right) - \frac{2}{15} {\frac{ \Hsq_{ab}}{\Hsq}\mathcal{O}^{ab}} \bigg]\,,
\\
&=&3+\frac{1}{H^2}\bigg\{ \frac{1}{6}\rho  -\frac{1}{3}\Lambda 
%-\frac{2}{15}\left[\frac{1}{6}\rho + \frac{1}{3}\Theta^2 -\frac{1}{3}\Lambda \right]\left[ \frac{\sigma_{\<ab\>}\sigma^{\<ab\>}}{H^2}   \right]
+\frac{3}{5}\sigma_{ab}\sigma^{ab}
%\\ \nonumber &&
+ \frac{6}{15} \left[ \frac{1}{6}\rho  -\frac{1}{3}\Lambda \right]\left[ \frac{\sigma_{\<ab\>}\sigma^{\<ab\>}}{H^2}   \right]
-\frac{4}{15} \frac{\sigma^{\<ab\>} E_{\<ab\>}}{H} 
\bigg\}\,.
\end{eqnarray}
Finally, we find that the monopole of the logarithm of the Hubble rate normalised luminosity distance is given by
\begin{eqnarray}\label{eq:monopolehatdL}
\averageA{\log_{10} \hat{d}_{L}(z,{\n})}  &=&\log_{10} \bigg\{z\bigg[1+ \frac{1}{2} \bigg[1-\frac{1}{H^2}\bigg(\frac{1}{6}\rho  -\frac{1}{3}\Lambda 
+\frac{3}{5}\sigma_{ab}\sigma^{ab}
+ \frac{6}{15} \left[ \frac{1}{6}\rho  -\frac{1}{3}\Lambda \right]\frac{\sigma_{\<ab\>}\sigma^{\<ab\>}}{H^2}  
\\ \nonumber &&
-\frac{4}{15} \frac{\sigma^{\<ab\>} E_{\<ab\>}}{H} 
\bigg)
\bigg]z+ \mathcal{O}(z^2)\bigg\}\,.
\end{eqnarray}
Putting everything back to equation \eqref{eq:mu_averaging}, we find that the monopole of the distance modulus is given by
\begin{eqnarray}
%\averageA{\mu}&=& \averageA{m}-\averageA{M}^{R} =25 -5  \log_{10}{H_0} + 5\averageA{ \log_{10}\left[  \hat{d}_{L}(z,{\n}) \right]}\,,\\
\averageA{\mu}&=&\averageA{m}-\averageA{M}^{R}=25 +5  \log_{10} \bigg\{ \frac{1}{{H}^{\mu}_0}\bigg[z\bigg(1+ \frac{1}{2} \big(1-{q}^{\mu} 
\big)z+ \mathcal{O}(z^2)\bigg)\bigg]
\bigg\}\,,
\end{eqnarray}
where the following terms were defined by comparing to the equivalent FLRW expression given in equation \eqref{eq:mudL} and \eqref{eq:mudL1}
\begin{eqnarray}
\averageA{M}^{R} &=& \averageA{M} + 5  \log_{10}  \left[1+\frac{1}{15}\frac{\sigma_{ab}\sigma^{ab}}{H^2}\bigg|_{z=0 }\right] \,,
\label{eq:Mrenorm}
\\
{H}^{\mu}_0&=&H_0\,,
\\
{q}^{\mu} &=& \frac{1}{H^2}\bigg(\frac{1}{6}\rho  -\frac{1}{3}\Lambda 
+\frac{3}{5}\sigma_{ab}\sigma^{ab}
+ \frac{2}{3} \left[ \frac{1}{6}\rho  -\frac{1}{3}\Lambda \right]\frac{\sigma_{\<ab\>}\sigma^{\<ab\>}}{H^2}  
-\frac{4}{15} \frac{\sigma^{\<ab\>} E_{\<ab\>}}{H} 
\bigg)\bigg|_{z=0}\,.
\end{eqnarray}
 This indicates that the best-fit absolute magnitude, $\averageA{M}^{R}$ of the SNIa includes the effect of tidal deformation of the local geometry.  Stated differently, the measurement of the Hubble rate via the local distance ladder accounts for the impact of tidal deformation on the luminosity distance. The Hubble rate determined using the local distance ladder with a properly calibrated absolute magnitude corresponds to the global volume expansion.

\subsection{Cosmological fitting problem:  intercept of the Hubble diagram}\label{sec:DMI}

On the other hand, the SH0ES collaboration uses the Cepheid variable stars to calibrate the absolute magnitude of the SNIa. The estimate of the Hubble rate is obtained from the constraint on the intercept of the magnitude-redshift relation with the luminosity distance given by the flat FLRW spacetime. Starting from equation \eqref{eq:mag-redshift-relation}, they re-write the distance modulus as follows \cite{Riess:2016jrr}
\begin{eqnarray}
m &=& M + 25 + 5 \log_{10} \bar{d}_{L} = -5a_b + 5 \log_{10} \hat{d}_{L} (z)\,,
\label{eq:mag-red-relation2}
\end{eqnarray}
where $a_b$ is the intercept of the of the Hubble diagram and $\hat{d}_{L}$ is the Hubble parameter normalised luminosity distance: 
\begin{eqnarray}\label{eq:intercept}
5 a_b&=& -\left( M_b+ 25 - 5 \log_{10} H_0\right)\,, \\
\hat{d}_{L}(z) &=& {H_0} d_{L}(z) =  z \left[ 1 + \frac{1}{2}(1-q_0) z + \mathcal{O}(z^2)\right]\,.
\label{eq:lum-red-relation}
\end{eqnarray}
%Here, $M_b$ is the standardized absolute luminosity of a type Ia supernovae.
 The Hubble rate is then determined from equation \eqref{eq:intercept}
\begin{eqnarray}\label{eq:FLRW_intercept}
\log_{10} H_0= \frac{ M_b+25  + 5a_b}{5}\,,
\end{eqnarray}
where $M_b$ is called standardizable absolute luminosity in \cite{Riess:2016jrr} but it does the same job as equation \eqref{eq:absMag} in CCHP
\begin{eqnarray}\label{eq:standardizable}
M_b = m_{b,\rm{SNIa}} -\mu_{0,\rm{ceph}} %= M -\mu_{0,\rm{ceph}}\,,
\end{eqnarray}
where, $\mu_{0,\rm{ceph}}$, is an independent distance modulus to cepheids and $ m_{b,\rm{SNIa}}$, is the apparent magnitude of a sub-sample of nearby SNIa that live in the same host as the cepheids. 
The SH0ES collaboration has since made use of different geometrical distance estimates such as the NGC 4258 obtained by modelling of the water masers in the nucleus of the galaxy orbit about its supermassive black hole~\cite{Reid:2019tiq}; Large Magellanic Cloud using eclipsing binary systems composed of late-type stars~\cite{2019Natur.567..200P}; Milky Way Cepheids using parallax methods~\cite{Riess:2018byc,2018ApJ...855..136R} to calibrate $M_b$.
%Similar to the CCHP,  this term emerges naturally when the effects of the tidal distortion of the local spacetime is taken into account.  The Hubble rate is obtained from equation \eqref{eq:intercept}
Even though the  SH0ES collaboration limits type Ia supernova sample to $z =0.023$ due to systematics associated with disentangling the peculiar velocities of the sources from the coherent Hubble flow, they have accurate information on the intercept 
\begin{eqnarray}
a_b &=& \log_{10} \bigg\{  cz \left[ 1 + \frac{1}{2}(1-q_0) z + \mathcal{O}(z^3)\right]\bigg\} - 0.2 M_b\,,
\label{eq:FLRWa}
%\\
%&=& \log_{10}cz+  \log_{10} \left[ 1 + \frac{1}{2}(1-q_0) z+ \mathcal{O}(z^3)\right] - 0.2 M
\\
&\approx&\log_{10}  cz  - 0.2 M_b\,,
\end{eqnarray}
Similarly, the SH0ES collaboration  puts uninformative prior on {$\bar{q}_{0} = -0.55$}~\cite{Riess:2006fw} to determine $a_b = 0.71273 \pm 0.00176$. With the constraint on $M_b$, they found that the Hubble rate is given by~\cite{Riess:2019cxk,Riess:2020fzl}
\begin{eqnarray}
H_{0} = 73.1 \pm 1.4 \,\,{\rm{kms}^{-1} } {\rm{Mpc}}^{-1}\,.
\end{eqnarray}
Note that the quoted error here are the error associated with the determination of  $a_b$ and $M_b$ added in quadrature~\cite{Riess:2016jrr}.

%\subsubsection{In the full spacetime}

We will now show how the approach taken by the SH0ES collaboration may be generalised to apply to arbitrary inhomogeneous models. The key ingredient in the generalised expression is the luminosity distance given in equation \eqref{eq:Glumdist}. Starting from equation \eqref{eq:mag-redshift-relation} and following similar steps that starts from equation \eqref{eq:mag-red-relation2} in the FLRW limit
\begin{eqnarray}
 \averageA{m}  &=&\averageA{M} -{5}\averageA{\log_{10}[K^cK^d\nabla_du_c]_o} +5 \averageA{ \log_{10}\left[  \hat{d}_{L}(z,{\n}) \right]}+25 \,,
 \\
 &=& -5\averageA{a_b} + 5\averageA{ \log_{10} \hat{d}_{L} (z,{\n})}\,,
 \label{eq:m_averaging_SHoES}
\end{eqnarray}
where $\hat{d}_{L} (z,{\n})$ is given in equation \eqref{eq:mag-red-relation3} and  the  generalised intercept is given by
\begin{eqnarray}\label{eq:average_intercept}
 \averageA{a_b({\n})} = -\frac{1}{5}\left( \averageA{M}+ 25 - 5 \averageA{\log_{10} \left[K^aK^b\nabla_a u_b\big|_{0}\right]}\right)\,.
\end{eqnarray}
Recall that the decomposition of $\averageA{K^aK^b\nabla_a u_b}$ is discussed between equation \eqref{eq:KKDu-decomp} to \eqref{eq:AverageKKDu}.  Putting this back into equation \eqref{eq:average_intercept}, gives
  \begin{eqnarray}\label{eq:average_intercept2}
 \averageA{a_b({\n})}  &=& -\frac{1}{5}\left(\averageA{M}^{R}+ 25 - 5\log_{10} H_0\right)\,,
\end{eqnarray}
where we have introduced a renormalised absolute magnitude defined in equation \eqref{eq:Mrenorm}.
Similarly, the Hubble rate depends on the monopole of the  intercept  and renormalised absolute luminosity according  to 
 \begin{eqnarray}\label{eq:average_intercept2}
\log_{10} H_0= \frac{ \averageA{M}^{R}+25  + 5\averageA{a_b}}{5}\,.
\end{eqnarray}
 $\averageA{a_b}$ is obtained independently by fitting to the intercept of the distance modulus of the SNIa peak magnitudes. The  intercept of the distance modulus constrains ${\averageA{M}^{R}}+ 5 \averageA{a_b}$, from where we find 
 \begin{eqnarray}\label{eq:averageA}
    \averageA{a_b({\n})} &=&\averageA{\log_{10} \hat{d}_{L}(z,{\n})}  - 0.2 \averageA{M}^R\,,
\\
 &\approx&\log_{10}\bigg\{cz\left[1+\frac{1}{2}\left[4-\averageA{\frac{K^cK^bK^a\nabla_a\nabla_bu_c}{\left(K^cK^b\nabla_cu_b\right)^2}}\right]_0z   + \mathcal{O}(z^2)\right]\bigg\}- 0.2\averageA{M}^R\,, 
 % \\ &\approx&\log_{10}  cz  - 0.2 \averageA{M}\,.
\\ &=&\log_{10} \bigg\{cz\bigg[1+ \frac{1}{2} \bigg[1-\frac{1}{H^2}\bigg(\frac{1}{6}\rho  -\frac{1}{3}\Lambda 
+\frac{3}{5}\sigma_{ab}\sigma^{ab}
+ \frac{6}{15} \left[ \frac{1}{6}\rho  -\frac{1}{3}\Lambda \right]\frac{\sigma_{\<ab\>}\sigma^{\<ab\>}}{H^2}  
\\ \nonumber &&
-\frac{4}{15} \frac{\sigma^{\<ab\>} E_{\<ab\>}}{H} 
\bigg)
\bigg]z+ \mathcal{O}(z^2)\bigg\}- 0.2 \averageA{M}^R\,,
\\
  &\approx&\log_{10}  cz  - 0.2 \averageA{M}^R\,.
\end{eqnarray}
where we made use of equation \eqref{eq:KKKuuKKu} in the second equality, in the third equality, we used equation \eqref{eq:monopolehatdL} and the very low redshift approximation is enough~\cite{Riess:2016jrr}.

Finally, we remark that the Hubble rate and the deceleration parameter obtained from fitting to  the monopole of the Alcock-Paczy\'{n}ski parameters for the  BAO signal in the two-point correlation function agree:
\begin{eqnarray}
{H_0^{d_L}} &=&{H_0^{d_A}} = H^{\alpha{\p}}_{0} \,,
\\
q^{d_L}_0 &=& q^{d_A}_0\,.
\end{eqnarray}
But they differ from the local cosmic distance ladder measurement determination
\begin{eqnarray}
H^{\mu_I}_{0} &=&{H}^{\mu}_{0} \,,
\\
q_{0}^{\mu_I} &=& q_{0}^{\mu}\,.
\end{eqnarray}
The Hubble rate from local cosmic distance measurement corresponds to volume expansion  without any contamination due to the tidal deformation of the local spacetime geometry.
The Hubble rate and the deceleration parameter determined from the local cosmic ladder and Alcock-Paczy\'{n}ski parameters of the BAO signal  differ according to
\begin{eqnarray}\label{eq:eff-Hubble}
{H_0^{d_A}}-H^{\mu}_{0}
&\approx& -\frac{2}{15}\frac{\sigma_{ab}\sigma^{ab}}{{H}}\bigg|_{z =0}+\mathcal{O}(\sigma^3)\,,
\\
\label{eq:qodif}
q^{d_A}_0  - q_{0}^{\mu} &\approx& -\frac{2}{15}   \frac{\sigma^{\<ab\>} E_{\<ab\>}}{H^2}\bigg|_{z=0}+\mathcal{O}(\sigma^3) .
\end{eqnarray}
It is possible to calibrate the SNIa using the BAO distance information (proper distance travelled by the acoustic waves from the initial time to the last scattering surface) instead of the nearby distance anchors~\cite{Aubourg:2014yra}. This approach usually involves the use of a model which does not include the tidal fields we discussed here in the calibration process~\cite{Camarena:2019rmj}, thus, we can associate the absolute magnitude of the SNIa determined through this process to $\averageA{M}$. On the other hand, the local cosmic distance ladder approach does not assume any model for distance(apart from the parallax formula for distance, flux inverse square law, period-luminosity relation for cepheids)  during the calibration process, hence, we can associate the absolute magnitude it determines to $\averageA{M}^R$. The difference is given by
\begin{eqnarray}
\averageA{M}^{R} - \averageA{M} &\approx&  \frac{1}{3\log10} \frac{ \sigma_{ab}\sigma^{ab}}{{H}^2}\bigg|_{z=0} +\mathcal{O}(\sigma^3)\,.
\end{eqnarray}

\section{The  perturbed FLRW spacetime}\label{sec:perturbation}

The results we have derived so far by fitting the FLRW model to an inhomogeneous model is general, they apply to any inhomogeneous cosmological model provided that anisotropies are small when compared to the monopole. In this section, we specialise to a Universe where the inhomogeneities could be described as small perturbations on top of a flat FLRW background spacetime. We work in conformal Newtonian gauge (our result is gauge-invariant): 
\begin{eqnarray}\label{eq:metric}
\d s^2 &=& \bar{a}^2\big[-\big[1 + 2\Phi \big]\d \eta^2 
 + \big(\big[1-2 \Phi\big]\delta_{ij} \big]\,,
\end{eqnarray}
where $\delta_{ij}$ is the metric of the Minkowski spacetime, $\bar{a}$ is the scale factor of the expanding background FLRW spacetime and $\Phi$ is the Newtonian  gravitational potential.   The components of the time-like four velocity in perturbation theory are given by
\begin{eqnarray}\label{eq:v0}
u^0&=&1 - \Phi +  \frac{3}{2} \Phi^2 -  \frac{1}{2}\Phi\two + \frac{1}{2}{\partial}_{i}v{\partial}^{i}v\,,\\
u^i&=&{\partial}^{i}v + \frac{1}{2}
v^{i}{}\two + \frac{1}{2} {\partial}^{i}v\two\,.
\label{eq:vi}
\end{eqnarray}
where $v^i$ is the peculiar velocity,  $\partial_i$ is the spatial derivative on the Minkowski spacetime in cartesian coordinates, and $v$ is the peculiar velocity potential. The indices with the small English alphabets denote the spatial component of the spacetime. {The green superscripts denote terms evaluated at second order in perturbation theory. }

%\subsection{Determinations of the Hubble rate in perturbation theory}

Using equations \eqref{eq:metric}, \eqref{eq:v0}  and \eqref{eq:vi} at leading  order in perturbation theory, the shear tensor, $\sigma^{ab}$  becomes
\begin{eqnarray}\label{eq:shearpert}
 {{\sigma}}_{ij}(\eta, {\x}) = \bar{a}\partial_{\<i} \partial_{j\>} v (\eta, {\x}) \,.
\end{eqnarray}
The ensemble average of the  shear scalar is given
\begin{eqnarray}\label{eq:sigmasq}
\frac{\sigma_{ab}\sigma^{ab}}{{H}^2}&=&%\frac{1}{{H}^2} \partial_{\<i} \partial_{j\>} v  \partial^{\<i} \partial^{j\>} v=
% (f(\eta))^2\partial_{\<i} \partial_{j\>\delta_{m}(\eta, {\x})\partial^{\<i} \partial^{j\>} \delta_{m}(\eta, {\x}) 
  \frac{2}{3}f^2(z) \sigma^2_R(z)\,,
\end{eqnarray}
where $f$ is the rate of growth of structures.
We made use of the continuity  to express the velocity potential and the in terms of the matter density contrast $\delta_m$ 
 \begin{eqnarray}\label{eq:continuityeqn}
 v({\k},\eta) &=& \frac{\HH}{k^2}f(\eta)\delta_m({\k},\eta)\,.
 \end{eqnarray}
In addition, we introduced the variance in matter density field
\begin{eqnarray}\label{eq:cosmic_variance}
 \sigma^2_R&=&\int_{0}^{k_{\rm{UV}}}   \frac{{\d} k}{2\pi^2}   \,\left[k^2 P_{\rm{m}}(k)  \right]
= \int_{0}^{\infty}  \frac{{\d} k}{2\pi^2}\,\left[(kW(k R))^2 P_{\rm{m}}(k) \right]\,.
\end{eqnarray}
Here, $P_m$ is the power spectrum of the matter density field. We have introduced a top-hat window function, $W$, instead of a UV dependent momentum integral. The scale, $R$, will be fixed shortly. 
 For the different determinations of the Hubble rate, we focused on the dominant terms only, perturbing the monopole of various definitions of the Hubble rate give
\begin{eqnarray}
H_0^{\mu} &=&H_0^{\mu_I}\simeq \bar{H}_{0}\,,
\\
H_0^{d_A}&=&{H_0^{d_L}} = H^{\alpha{\p}}_{0} \simeq\bar{H}_o\left[1 -\frac{4}{45}f^2(z) \sigma^2_{R}(z)\bigg|_{0}\right]\,.
\end{eqnarray}
Note that $\Theta$ is well described by the FLRW background spacetime to  an accuracy better than $0.1\%$ \cite{Umeh:2010pr}, thus we neglect its perturbations and approximate $H_0^{\mu}$ and $H_0^{\mu_I}$ with the FLRW background prediction, $\bar{H}_{0}$. The relationship between the rest of Hubble rate determinations and $H_0^{\mu_I}$ is given
\begin{eqnarray}\label{eq:Hubblediscrepant}
\frac{H^{\alpha{\p}}_{0}}{H_0^{\mu_I} }=1  -\frac{4}{45}f^2(z) \sigma^2_{R}(z)\bigg|_{0}\,.
\end{eqnarray}
Therefore, the BAO determination of the Hubble rate differs from the local measurements by a factor of $1  -{4}f^2(0) \sigma^2_{R}/{45}$.
%Notice that the BAO  and CCHP  best-fit values, i.e ${H_0^{d_L}} = H^{\alpha{\p}}_{0}$ agree.  In practice, there is also no statistically significant difference between then
%~\cite{Freedman:2021ahq}. %We discuss this further in section \ref{eq:resultsanddiscussion}.
%\subsection{Determinations of the deceleration parameter in perturbation}
%
The electric part of the Weyl tensor, $E_{ab}$, at leading  order in perturbation theory is given by 
\begin{eqnarray}
 {E}_{ij}(\eta, {\x}) &=&  \partial_{\<i}\partial_{j\>} \Phi(\eta, {\x}) \,.
\end{eqnarray}
We use the Poisson equation  to express  $\Phi$ in terms  $\delta_m$
\begin{eqnarray}\label{eq:Poissoneqn}
 \Phi({\k},\eta)& =& -\frac{3}{2}\Omega_m(z) \left(\frac{\HH}{k}\right)^2\delta_m({\k},\eta)\,,
\end{eqnarray}
where $\Omega_m$ is the  matter-energy density parameter.
  Using equations \eqref{eq:continuityeqn}  and \eqref{eq:Poissoneqn}  we find that
  \begin{eqnarray}
 \frac{\sigma^{ab}E_{ab}} {{H}^3} &=&%\frac{1}{{H}^3} \partial^{\<i} \partial^{j\>} v \partial_{\<i}\partial_{j\>} \Phi=-\frac{3}{2} \Omega_{m}f(\eta){D}_{ij}  \delta_{m}(\eta, {\x})  {D}^{ij}  \delta_{m}(\eta, {\x})=
   - \Omega_{m}f(z) \sigma^2_R(z)\,.
 \label{eq:sigmaEab}
  \end{eqnarray}
Expanding the first two terms in $q_0 = \left( \frac{1}{6}\rho -\frac{1}{3}\Lambda + \sigma^{ab}\sigma_{ab}\right) /{H}^2_{0}$ up to second order in perturbation theory gives
\begin{eqnarray}
 \frac{1}{H^2_{0}} \left(\frac{1}{6} \frac{1}{\rho} -\frac{1}{3} \Lambda \right)\bigg|_{z_o}  &\approx& \bar{q}_{0} - \frac{1}{3H_0} \Omega_m\delta_{m}  \partial^2 v + \frac{1}{3H^2_0}\bar{q}_{0}(\partial^2v)^2\,,
 \\
&\approx&  \bar{q}_{0} +\frac{1}{3} \Omega_mf(0)\sigma_{R}^2 +\frac{1}{3} \bar{q}_{0}f^2(0) \sigma_{R}^2\,,
\label{eq:qopartpert}
\end{eqnarray}
The product of $q_0$ and shear scalar in  perturbation theory  is given by
\begin{eqnarray}
 \frac{1}{H^2_{0}} \left(\frac{1}{6} \frac{1}{\rho} -\frac{1}{3} \Lambda \right)\bigg|_{z_o} \frac{\sigma_{ab}\sigma^{ab}}{H^2} \bigg|_{z_o} = \frac{2}{3} \bar{q}_{0}f^2 (0)\sigma^2_R\,,
\end{eqnarray}
where $\bar{q}_{0}$ is the deceleration parameter on the background FLRW spacetime with  dust and Cosmological constant
\begin{eqnarray}
\bar{q}_{0} &=&  \frac{\Omega_m}{2}  - \Omega_{\Lambda}  = - 1 + \frac{3}{2} \Omega_{m}\,,
\end{eqnarray}
where  $ \Omega_{\Lambda}$ is the energy density due to the Cosmological constant.
In the  second equality, we made use of  the  Friedmann equation
$
\Omega_m + \Omega_{\Lambda}  = 1\,,
$  to express $\bar{q}_{0} $ in terms of $\Omega_m$ only.  The matter density and the cosmological constant today are defined in terms of these parameters  as  $\bar{\rho}_{0} = 3 \HH^2_{0} \Omega_{m0}$ and $\Lambda = 3 \HH^2_{0} \Omega_{\Lambda 0}$.
Using equation \eqref{eq:qopartpert},  $q_0$ perturbed up to second order is given by
\begin{eqnarray}
\frac{1}{{H}^2_{0}} \left( \frac{1}{6}\rho -\frac{1}{3}\Lambda + \sigma^{ab}\sigma_{ab}\right)  \approx \bar{q}_{0} +\frac{1}{3} \Omega_mf(0)\sigma_{R}^2 +\frac{1}{3} \bar{q}_{0}f^2(0) \sigma_{R}^2 +\frac{2}{3}(f^2(0)) \sigma^2_R(z)
\end{eqnarray}
%On the background, this term simplifies as 
%\begin{eqnarray}
%\frac{1}{H^2} \bigg(\frac{1}{6}\rho -\frac{1}{3}\Lambda  - \frac{1}{3}H^2
%\bigg)\bigg|_{z=0}= \frac{1}{H^2} \bigg(\frac{1}{6}\rho -\frac{1}{3}\Lambda  
%\bigg)- \frac{1}{3} = q^{\rm{FLRW}}_{0}- \frac{1}{3} \,.
%\end{eqnarray}
Finally,  the  deceleration parameter for both cases becomes
\begin{eqnarray}
q_0^{d_A}=q_0^{d_L}&\simeq&\bar{q}_{0}+ \frac{11}{15} \Omega_m  f(0) \sigma_R^2 
+ \left( \frac{7}{9} \bar{q}_{0} + \frac{2}{5} \right) f^2(0)\sigma^2_{R}
 \,,
\label{eq:decerlationparamtodaydAsig}
\\
q^{\mu}_0=q^{\mu_I}_0&\simeq&\bar{q}_{0} +\frac{3}{5}  \Omega_mf(0)\sigma_{R}^2 +  \left(\frac{7}{9}\bar{q}_{0}+ \frac{2}{5} \right)f^2(0) \sigma^2_R\,.
\label{eq:decerlationparamtodayDMIsig}
\end{eqnarray}
The exact magnitude of the difference between the predications of the  effective Hubble rate and deceleration parameter from different ways of fitting  an FLRW model to a lumpy Universe depends on $\sigma^2_R$.  We discuss how we  handle the dependence on  $\sigma^2_R$ in section \ref{sec:causalhorizon}.

\subsection{Causal horizon and the expanding regions of spacetime }\label{sec:causalhorizon}

%\subsection{Existence of casual horizon for the time-like geodesics }\label{sec:causalhorizon}

In cosmology, the well-known causal limits are determined by the dynamics on the null cone, for example, the particle horizon,  
which indicates the maximum distance light from particles could have travelled to the observer in the age of the Universe. 
 Ellis and Stoeger argued in \cite{Ellis:2010fr} that there exists a causal horizon that is determined not by the dynamics on the light cone but by the dynamics of our time-like geodesic. It is given by the boundary of the comoving region that has contributed most significantly to the dynamics of our local environment.
The dominant interaction within this region are not mediated by massless particles (i.e the vector and tensor perturbations on an FLRW background spacetime have negligible impact on the dynamics within this region), instead, they are mediated by massive particles which travel at very low speeds relative to the cosmic rest frame. It is these differences in speed that cause our local environment to decouple from the Hubble expansion because it cannot keep up the cosmic rest frame~\cite{Ellis:2010fr,Ellis:2020kry}. 
 \begin{figure}[h]
%\centering 
\includegraphics[width=85mm,height=75mm] {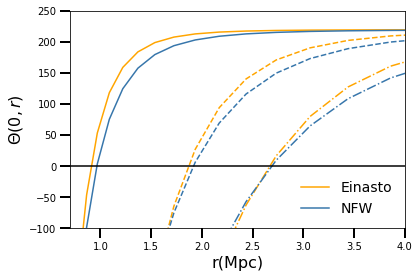}
\includegraphics[width=85mm,height=75mm] {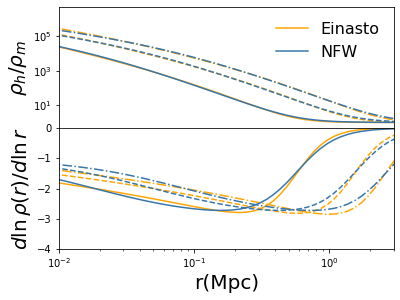}
\caption{\label{fig:theta_crit}  %
The horizontal line in the left panel shows the zero-velocity(expansion) surface: $\Theta(0,R{\n}) =0$, i.e the transition radius where the spacetime go from expansion to contraction. The exact position of $R$ depends on the halo mass. The three curves; {thick, dashed and dash-dotted lines} correspond to the following halo masses $M=[10^{11},10^{12},10^{13}] M_{\odot}$ respectively.  
On the right panel, we compute the gradient of the halo density only for the same halo masses shown in the left panel. 
 The splashback radius for a given halo mass is the position of the least gradient. The splashback radius is always less than the radius of the zero-velocity surface. 
In both cases, we fixed the halo concentration at 5. 
}
\end{figure}
We determine the causal horizon as the comoving radius where divergence of our 4-velocity vanish. This is also called the zero-velocity surface~\cite{Karachentsev:2008st}.  At the leading order in cosmological perturbation theory, this is given by
\begin{eqnarray}\label{eq:expansion}
\Theta \simeq 3 H_0 + \partial_{i}\partial^i v\big|_{z=0}% = 3 H _0- \delta'_{m}\big|_{z=0}
=3 {H}_0 +  \frac{c}{r} \frac{\d \ln \rho}{\d \ln r} \,.
\end{eqnarray}
%Again, the vector and tensor perturbations on an FLRW background spacetime are mediated by massless particles therefore, have would a negligible impact on equation \eqref{eq:expansion}.
 In the second equality, we made use of the continuity equation: $\delta'_{m} = -\partial_{i}\partial^i v$ with
$\delta_{m}  \equiv {\delta \rho}/{\bar{\rho}} = ({\rho - \bar{\rho}})/{\bar{\rho}}\,$  and use the chain rule to express the conformal time derivative of $\delta_{m} $ in terms of  radial derivative of $\rho$
%the conformal time derivative of $\delta_{m} $ as to radial derivative
\begin{eqnarray}\label{eq:deltampr}
\delta'_{m} = \frac{\d \delta_{m}}{\d \eta} =\frac{1}{\bar{\rho}} \frac{\partial r}{\partial \eta} \frac{ \partial \rho}{\partial r}  \approx - \frac{1}{\bar{\rho}} \frac{ \partial \rho}{\partial r}  
\approx - \frac{c}{r} \frac{\d \ln \rho}{\d \ln r}\,.
\end{eqnarray}
The radius of the zero-velocity surface $\Theta(0,{r})  =0$ is given by
\begin{eqnarray}
R_{0} = - \frac{ c}{ 3 H_0} \frac{\d \ln \rho}{\d \ln r}. %\approx - \frac{ 1}{ 3 } \frac{\d \ln \rho}{\d \ln r}
\end{eqnarray}
The spacetime in the  region $r<R_{0}$ are not expanding and it defines the local sphere of influence with respect to the observer. 
We estimate equation \eqref{eq:expansion} using a dark matter halo model with the Einasto~\cite{1989AandA...223...89E} and NFW profiles~\cite{Navarro:1995iw} with the outer profile given by the mean matter density~\cite{Diemer:2017bwl}. The result is shown in figure \ref{fig:theta_crit}. {$R_{0}$ is obtained from  figure \ref{fig:theta_crit} as the value of $r$ where the curves intersect the $\Theta(0,r) =0$ horizontal line.  }
We find that $R_{0}$ is dependent on the mass of the host halo. The larger the halo mass, the higher the causal horizon. One might ask, how is $R_{0}$ related to the virial radius, $r_{\Delta_c}$($\Delta_{c}$ is over-density constant). The mass contained with $r_{\Delta_c}$ (virial mass) is the mass of a gravitationally bound astrophysical system, assuming the virial theorem holds. Recent studies have shown that $r_{\Delta_c}$ does not correspond to the physical boundary of a gravitationally bound astrophysical system. The reason is that the mass contained within $r_{\Delta_c}$ is subject to pseudo-evolution~\cite{More:2015ufa,Umeh:2021xqm}.  
And there are physical processes that are known to redistribute sub-structures formed during collapse from small to large radii greater than $r_{\Delta_c}$~\cite{Kazantzidis:2005su,Valluri:2006vi,Carucci:2014ema}. For these reasons, the splashback radius, $r_{\rm{sp}}$, was introduced as a radius that includes all matter that orbits a halo~\cite{Diemer:2014xya,More:2015ufa}. The splashback radius of haloes of different masses is shown in the right panel of figure \ref{fig:theta_crit}.  
We find that $R_{0}$ is different from the splashback radius of the host halo. In fact from figure \ref{fig:theta_crit}, we find that $R_{0}$ is always greater than the $r_{\rm{sp}}$ at any of the given halo mass we considered.  There are observational constraints for the radius of the zero-velocity surface for our local group $R_0\sim(0.95- 1.05) {\rm{Mpc}} $~\cite{1999AandARv...9..273V,Li:2007eg,Karachentsev:2008st} {and this is the value we use for the rest of the analysis}.

%\blue{Am important paper for the radius of the  zero velocity surface \cite{Karachentsev:2008st}}

\subsection{The cosmological  tensions: Hubble rate, absolute magnitude and the deceleration parameter}

\begin{figure}[h]
%\centering 
\includegraphics[width=80mm,height=65mm] {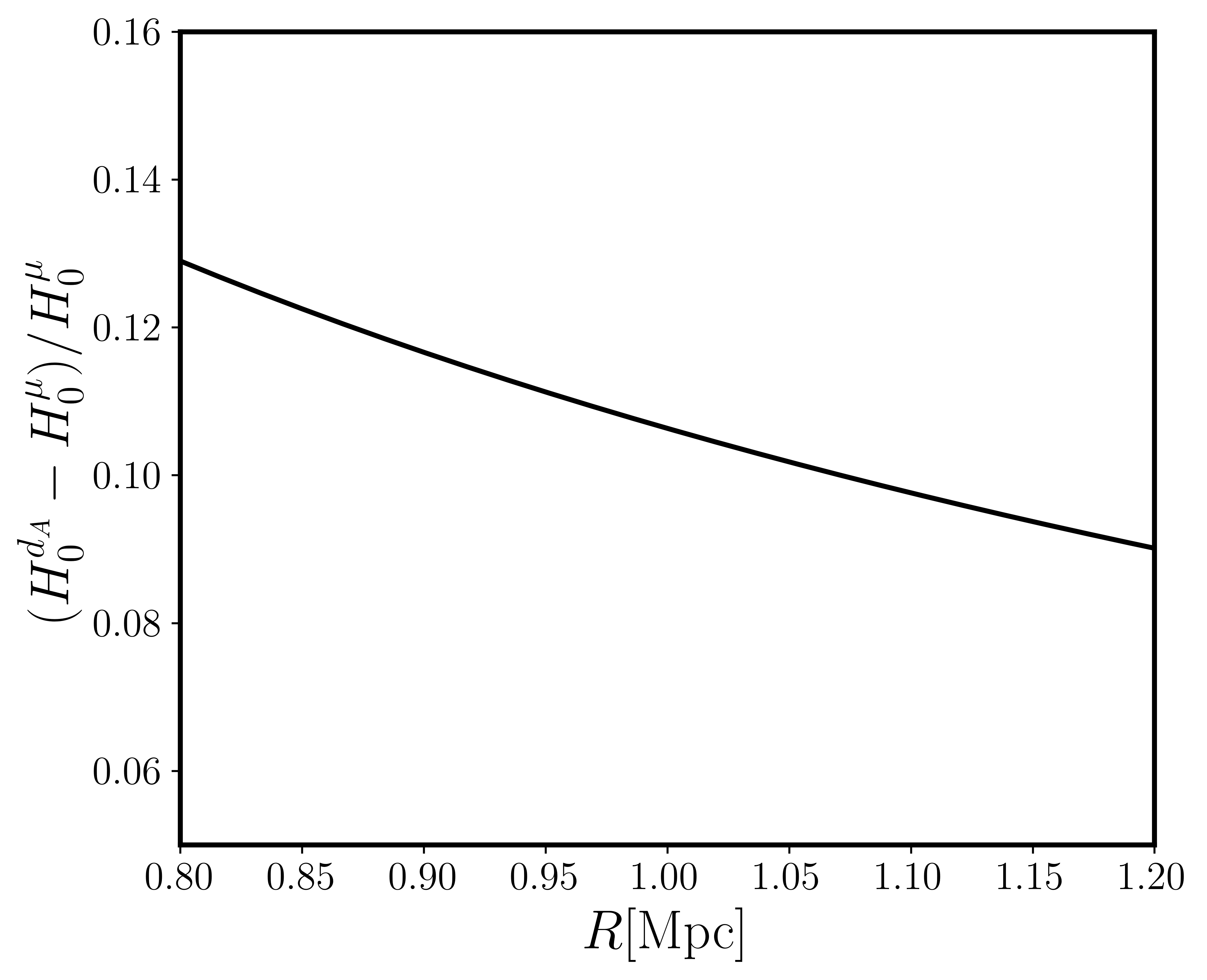}
\includegraphics[width=80mm,height=65mm] {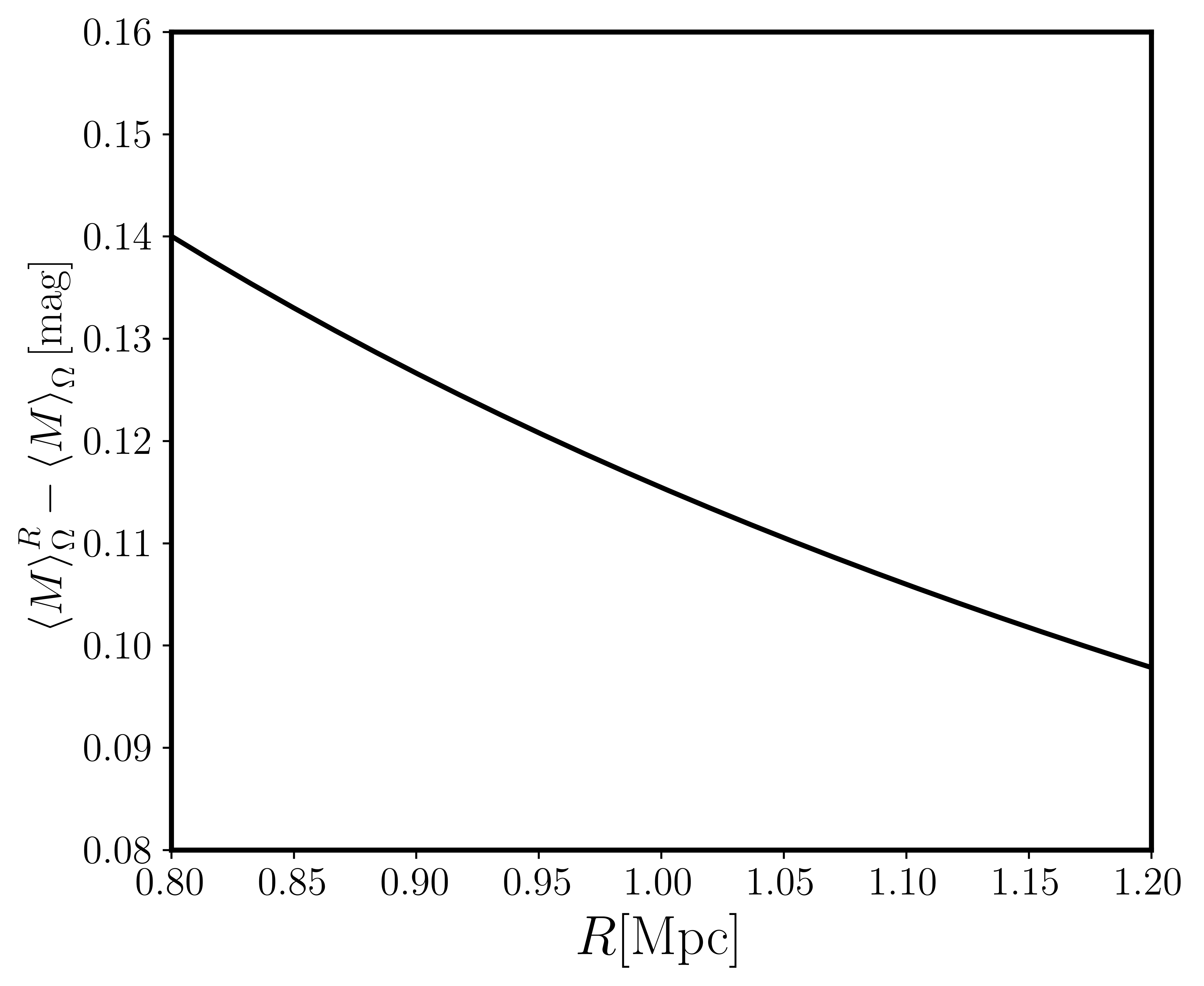}
\caption{\label{fig:Hubbleratet} Left panel: We show the fractional difference between the Hubble rates determined using the local measurement of the peak apparent magnitude of SNIa and the interpretation of the Alcock-Paczy\'{n}ski parameters in terms of the FLRW spacetime. 
Right panel: We show the difference between the renormalised absolute magnitude due to the impact of the tidal deformations on the distance to the anchor and the intrinsic absolute magnitude. 
We made use of the linear theory matter power spectrum to evaluate these.
 }
\end{figure}

 Now that we have determined the minimum length scale participating in the cosmic expansion, it is straightforward to calculate the Hubble discrepant (i.e equation \eqref{eq:Hubblediscrepant}).  
The minimum length scale corresponds to the distance where the reference flux or the absolute magnitudes of sources in the Hubble flow are calibrated. {This makes sense because cosmic evolutionary effect does  not  contribute to the absolute magnitude}. Setting the smoothing scale to the radius of the zero-velocity surface $R=R_0$. 
The full result is shown in figure \ref{fig:Hubbleratet}. We find about $\sim (9-12)\%$ difference in the determination of the Hubble rate between the local measurements using the cosmic distance ladder and the BAO determinations.  

%{\cite{Adler:2019fnp} studied the impact of a scale invariant but frame dependent dark energy on the local value of the Hubble constant. They showed that it could explain the SHoES result but it gives large value of the Hubble constant for the BAO contrary to observation.
%\cite{Capozziello:2020nyq} The uncertainty on measurements, \cite{Zumalacarregui:2020cjh}
% Summary of attempts }

%\subsection{The Supernova absolute magnitude tension}
Although, the cosmological interpretation of the peak magnitude of the SNIa by the SH0ES and CCHP groups was done using a smooth luminosity distance,  % just as in the case of the BAO Alcock-Paczy\'{n}ski parameters. 
Both groups differ on the calibration of the supernova absolute magnitude. The SH0ES  build a local cosmic distance ladder using distance measurements to a set of anchors to infer the absolute magnitude of the SNIa. Other than using the standard formula for the parallax for the milky-way cepheids or the inverse square law for flux for the detached eclipsing binaries, or Leavitt's law for cepheids~\cite{1930AnHar..85..143L}, etc, the process of calibrating the absolute magnitude of the SNIa is cosmological model-independent. Since these set of anchors live nearby, the calibration of the local cosmic distance ladder includes the impact of the tidal deformations on the local spacetime that we discuss here~\cite{Umeh:2022hab}. 
We showed how the impact of the tidal deformation results in a renormalisation of the absolute magnitude in subsection \ref{sec:distancemod}. 
Calibrating the SNIa absolute magnitude using the proper distance acoustic waves could have travelled from the origin of the Universe to the surface of the last scattering as an anchor, requires that we assume a particular model based on the FLRW spacetime. This model does not include the impact of the tidal field on distance measurement~\cite{Aubourg:2014yra}.

The supernova absolute magnitude tension refers to the disparity in the determination of the supernova absolute magnitude between the inverse distance ladder method with the sound horizon scale at the surface of the last scattering as an anchor and local cosmic distance ladder with the anchors in the nearby Universe~\cite{Camarena:2019rmj,Camarena:2021jlr}. Using the inverse distance method on Pantheon supernova peak magnitudes,~\cite{Efstathiou:2021ocp} found the absolute magnitude to be
\begin{eqnarray}\label{eq:P18}
M^{\rm{P18}} = - 19.387\pm 0.021\,\, {\rm{mag}} \,.
\end{eqnarray}
For this same supernova sample, but using the SH0ES Cepheid photometry for the geometric distance estimates, ~\cite{Efstathiou:2021ocp} finds the absolute magnitude to be
\begin{eqnarray}\label{eq:E21}
M^{\rm{E21}} = -19.214\pm 0.037\,\, {\rm{mag}} \,.
\end{eqnarray}
The difference between equation \eqref{eq:P18} and \eqref{eq:E21} is given by
\begin{eqnarray}
M^{\rm{E21}} - M^{\rm{P18}} = 0.173 \pm 0.04 \,\, {\rm{mag}}\,.
\end{eqnarray}
We can relate equation \eqref{eq:P18} to $ \averageA{M}$ and equation \eqref{eq:E21} to $\averageA{M}^{R} $ so that the difference between them is given by equation \eqref{eq:Mrenorm}
\begin{eqnarray}\label{eq:Mdiff}
\averageA{M}^{R} - \averageA{M} &=&   \frac{2}{9\log10}f^2(0) \sigma^2_{R}\,.
\end{eqnarray}
 We show the result of calculating equation \eqref{eq:Mdiff} in the right panel of figure \ref{fig:Hubbleratet}.  With $R=D= 1 {\rm{Mpc}}$, we find a diffrenece of about $0.12~ {\rm{mag}}$

In addition to the Hubble rate and the absolute magnitude, we also derived the expression for the deceleration parameter by fitting the FLRW model to the monopole of an inhomogeneous model of the area distance, distance modulus, etc. In this case, we find that both the local measurements and the BAO measurements (Alcock-Paczy\'{n}ski parameters) give deceleration parameter that differ substantially from the prediction of an FLRW model
\begin{eqnarray}
q_{0}^{d_A/\mu} - \bar{q}_{0} = \Delta q^{d_A/\mu}_{0}\,,
\end{eqnarray}
where $\Delta q^{d_A/\mu}_0$ depends on the scalar invariant of a combination of shear tensor and the electric part of the Weyl tensor, i.e. ${\sigma^{\<ab\>} E_{\<ab\>}}/{H^2}\big|_{z=0}$ and $\sigma_{ab}\sigma^{ab}/H^2\big|_{z=0}$. The corresponding cosmological perturbation limit of both expressions are given in equations \eqref{eq:decerlationparamtodaydAsig} and \eqref{eq:decerlationparamtodayDMIsig}.
Again the local cosmic ladder and the Alcock-Paczy\'{n}ski parameters determinations of the deceleration parameter differ only in the coefficient of this term ${\sigma^{\<ab\>}E_{\<ab\>}}/{H^2}\big|_{z=0}$(see equation \eqref{eq:qodif}). 
%The coefficient of this term in $q_0^{d_A}$ is about two-thirds of $q_0^{\mu}$.
${\sigma^{\<ab\>}E_{\<ab\>}}/{H^2}\big|_{z=0}$ is related to the spatial curvature of the local domain.
% we discuss its physical significance in~\cite{Umeh:2021c}.
%
 \begin{figure}[h]
%\centering 
\includegraphics[width=100mm,height=75mm] {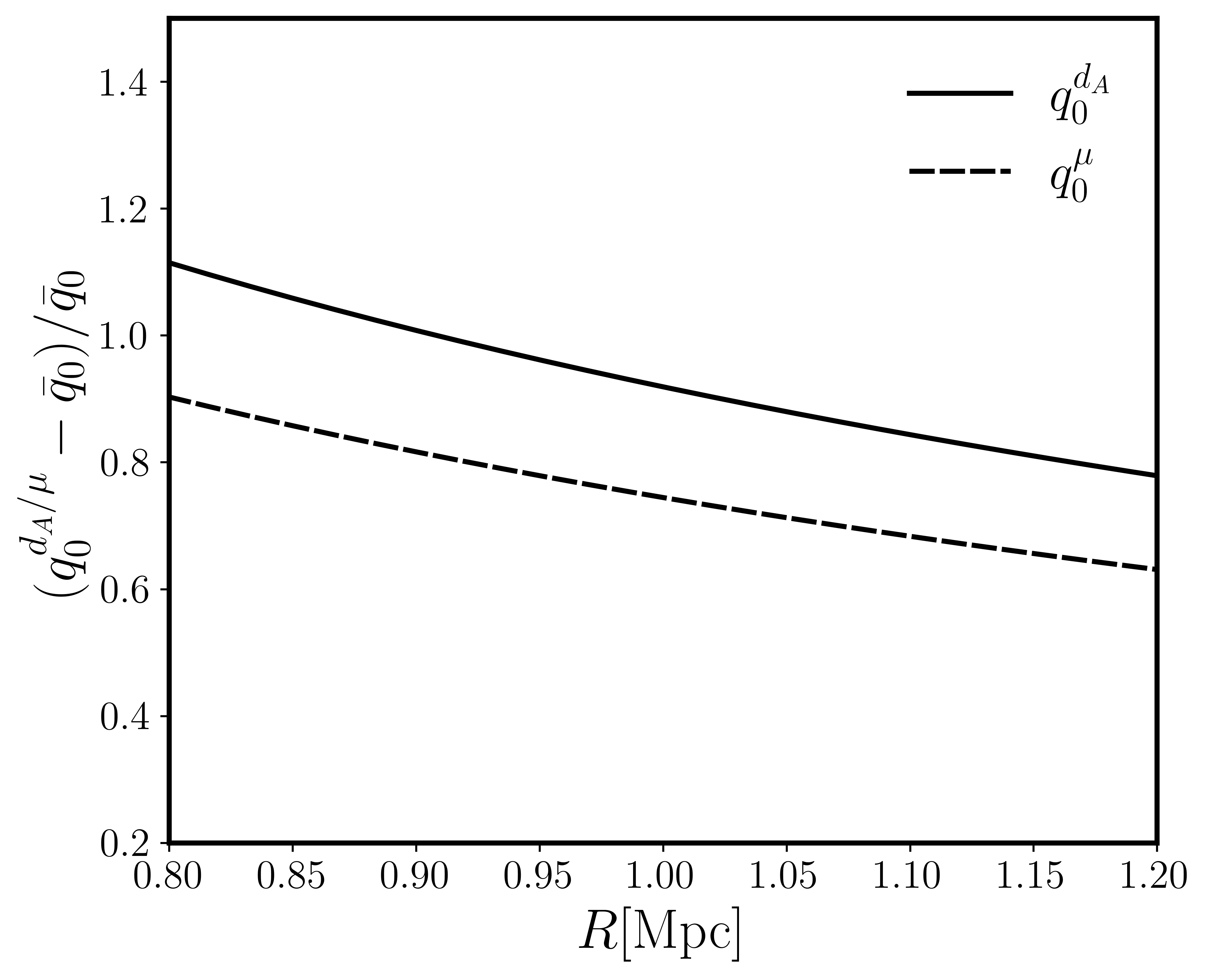}
\caption{\label{fig:decelerateratet}  We show the fractional difference between the deceleration parameters derived in equations \eqref{eq:decerlationparamtodaydAsig}(thick lines) and \eqref{eq:decerlationparamtodayDMIsig} (dashed lines) and the FLRW predictions. 
%Again, we made use of  the halo fit given in CAMB (blue curve)  and the linear matter power spectrum (orange curve) from CAMB as well~\cite{Lewis:1999bs}. 
%Right panel: Relative contribution to the total energy budget of the Universe assuming a flat spatial geometry. The tidal energy density parameter was estimated using equation \eqref{eq:sigmasq}. The vertical lines indicates where the energy density of the cosmological constant equals that of the matter density.
 }
\end{figure}

We show in figure \ref{fig:decelerateratet}, the  fractional difference between the deceleration parameters derived in equations \eqref{eq:decerlationparamtodaydAsig} and \eqref{eq:decerlationparamtodayDMIsig}  and the FLRW predictions. 
We found over a 100\% difference between the deceleration parameter from the local measurement interpreted based on an FLRW model and Planck prediction.  This finding is consistent with the recent measurements of the deceleration parameter using  the Pantheon supernova samples  and  the cosmographical expression for the luminosity distance in FLRW limit~\cite{Camarena:2019moy,Camarena:2021jlr}.  { However,  it differs from the measurement of the deceleration parameter by the SH0ES collaboration~\cite{Riess:2021jrx}. Note that SH0ES collaboration made use of the full expression for the luminosity distance based on the  background FLRW spacetime and not its cosmographical approximation~\cite{Riess:2006fw}. 
{Given that \cite{Riess:2021jrx} made use of the data set used by \cite{Camarena:2019moy,Camarena:2021jlr} but supplemented with another data set which is composed of high redshift SNIa, It is most likely that the reason for this disparity is related to the breakdown of Taylor series expansion in the presence of structures as reported in~\cite{Umeh:2022hab}.  %Essentially, this is indicating a breakdown of Taylor series expansion of the luminosity distance at second order in redshift.
 }}
%We point out that any unbiased determination of the deceleration parameter today, we must show clearly how it handles the impact of the tidal fields (${\sigma^{\<ab\>} E_{\<ab\>}}/{H^2}\big|_{z=0}$ and $\sigma_{ab}\sigma^{ab}/H^2\big|_{z=0}$) on the local spacetime. 

%t{ The plot in the left panel is subject to $\Omega_m + \Omega_{\Lambda}  = 1\,,$ }
% between the Planck prediction for q_{0}^{d_A/\mu} 

The Hubble tension is usually described as a discrepancy between the early and late-time Universe determination of the Hubble rate~\cite{DiValentino:2021izs}. We have shown that it {could be a consequence of how the distance measurements in the neighbourhood of the observer are interpreted within the $\Lambda$CDM model}. The key factor is that the local spacetime around the observer is tidally deformed,  the  background FLRW spacetime does not capture this effect. We showed that  the model of {area/luminosity distance that includes the effects of inhomogeneities through the perturbation of the FLRW spacetime would be able to explain the Hubble tension without any need of invoking any exotic evolving dark energy~\cite{Dainotti:2021pqg,Niedermann:2021vgd} (see \cite{Knox:2019rjx,Abdalla:2022yfr} for a list of all possible models within this framework), frame-dependent dark energy~\cite{Adler:2019fnp} (this approach explains the SH0ES result but gives  a large value of the Hubble constant from the BAO analysis), quantum measurement uncertainties~\cite{Capozziello:2020nyq}, evolving gravitational constant~\cite{Benevento:2022cql}, modification of gravity~\cite{Zumalacarregui:2020cjh,Dainotti:2022bzg}. One unique thing about all these approaches except the frame dependent dark energy which breaks 4D diffeomorphism invariance but retains 3D coordinate invariance is that they assume that the FLRW background spacetime is valid on all scales and at all times. This is contrary to the findings in \cite{Umeh:2022hab} which showed that the FLRW background spacetime in the presence of structures break down at about 1 Mpc. This scale is consistent with the scale where caustics or conjugate points appears~\cite{Pichon:1999tk}.  Appearance of the conjugate points is a unique indicator for the breakdown of the coordinate system~\cite{Witten:2019qhl}.}

 {Furthermore, we showed that the interpretation of the BAO measurement (i.e Alcock-Paczy\'{n}ski parameters) using the area distance based on the background FLRW spacetime infers a wrong $H_0$ because the symmetries of background FLRW spacetime does not allow the contribution of the electric part of the Weyl tensor or the effect of the tidal deformations around the observer.
If the 
Alcock-Paczy\'{n}ski parameters are consistently interpreted or evolved from high redshift to low redshifts using a model of the area distance that includes the impact of the tidal field  or the area distance derived from a perturbed FLRW spacetime as described in \cite{Umeh:2022hab},  the tensions in the Hubble parameter would not arise.} 

{Of course, we are not the first to claim that the effects of inhomogeneities could resolve the Hubble tension. In fact it was argued in \cite{Lombriser:2019ahl} that the large scale outflows due to the presence of  local voids could be the cause of the Hubble tension and it could potentially can explain the late time cosmic acceleration by dark energy~\cite{Kasai:2019yqn}.   The authors defined a spatial averaging procedure following the Buchert averaging formalism~\cite{Buchert:1999er}.  They found that an average under density of about $\<\delta\> = -0.3$ is enough to explain the discrepant Hubble rate.  A more detailed study of this has shown that this does work~\cite{Kenworthy:2019qwq}. Secondly, the authors did not compute the average values of what is observed rather, they calculated the average density on the spatial hypersurface. An observer does not have access to the entire hypersurface rather it has access only  to the screen space (2D surface) due to restriction imposed by the constant speed of light~\cite{Buchert:2022zaa}. Similarly, it has been conjectured that anisotropies in the  super-horizon  perturbations could explain the Hubble tension.  This approach has some fine-tuning issues to resolve.  This simplest model of this has three free parameters that have to be chosen precisely for it to work~\cite{Tiwari:2021ikr}. }

%approach developed by Kristian and Sachs \cite{1966ApJ...143..379K} to obtain 

%%%%%%%%%%%%%%%%%%%%%%%%%%%%%%%%%%%%%%%%%%%%%%%%%%%%%%%%%%%%%%%%%
\section{Conclusion}\label{sec:conc}

We have deployed the cosmological fitting approach introduced by Ellis and Stoeger~\cite{Ellis:1984bqf,Ellis:1987zz} to study how well the FLRW spacetime fits an inhomogeneous Universe on average.  The FLRW model could be characterised by a set of free parameters such as the Hubble rate, deceleration parameter, etc, we show how to obtain the best-fit Hubble rate and the deceleration parameter given generalised inhomogeneous models of the following observables: Alcock-Paczy\'{n}ski parameters, magnitude-redshift relation, the intercept of the distance modules. 

We made use of the results of the low redshifts Taylor series expansion of the generalised inhomogeneous expression of the area distance and luminosity distance developed by Kristian and Sachs in 1966 \cite{1966ApJ...143..379K}. Using the $1+3$ covariant decomposition formalism, we decomposed these expressions in terms of irreducible observables with respect to a geodesic observer. In this limit, the only key observables are the rate of expansion (a scalar) and the rate of shear deformation tensor. The rate of expansion scalar describes the rate at which nearby geodesics expands/contract with respect to a geodesic observer. It corresponds to the Hubble rate in the FLRW limit. The shear tensor, on the other hand, describes the rate of change of the spacetime deformation in the neighbourhood of the observer~\cite{Ellis:1998ct}. It vanishes in the FLRW limit.

Using these tools, we derived the generalised inhomogeneous equations for the Alcock-Paczy\'{n}ski parameters. The Alcock-Paczy\'{n}ski parameters constrain the imprints of the Baryon Acoustic Oscillation in the galaxy distribution through the N-point correlation function. These parameters are usually interpreted in terms of the background FLRW model. We showed how to generalise the two components of the Alcock-Paczy\'{n}ski parameters to an inhomogeneous spacetime model. By comparing the monopole of the generalised radial component to the corresponding FLRW counter-part, we find that the inferred Hubble rate is biased by the impact of the tidal deformation tensor at the observer location. The orthogonal part is proportional to the area distance, hence we compare the monopole of the generalised area distance to the corresponding FLRW limit. At the leading order in redshift, we find that the Hubble rate today is biased as well by the scalar invariant of the shear tensor. At second order in redshift, we obtain the generalised expression for the deceleration parameter which is also biased by the scalar invariant of the shear tensor and the product of the shear tensor and the electric part of the Weyl tensor.

Furthermore, we considered the magnitude-redshift relation. This is the key observable from which the determination of the Hubble rate is made from the peak magnitude of the type IA supernovae. We showed that the tidal deformation tensor that biases the determination of the Hubble rate from the monopole of the Alcock-Paczy\'{n}ski parameters, impacts the measurement of the supernova absolute magnitude instead. The Hubble rate, in this case, is not biased, it corresponds exactly to the rate of expansion scalar which describes the rate of volume expansion/contraction. The deceleration parameter, however, is biased by the same terms as in the case of the orthogonal component of the Alcock-Paczy\'{n}ski parameter.

We quantified these terms within the cosmological perturbation theory assuming Gaussian and adiabatic initial conditions. We showed that the expectation value of these biasing terms is proportional to the smoothing scale-dependent variance in the matter density field. We argued that the most physically motivated smoothing scale corresponds to the comoving radius (causal horizon) where the spacetime region goes from contracting to expanding phase. We use the halo model to quantify this exactly. We find that it is slightly greater than the splashback radius of the host halo. 
Setting the mass of our local group to about$\sim(10^{11}-10^{12})M_{\odot}$, we find the smoothing scale in the range of $R\sim(0.8- 1.2) {\rm{Mpc}}$. This leads to about$\sim (8-12)\%$ fractional difference in the determination of the Hubble rate between the local measurements and the Alcock-Paczy\'{n}ski parameters.
With the same range of values for the smoothing scales, we find that it explains the supernova absolute magnitude tension~\cite{Camarena:2021jlr,Efstathiou:2021ocp} and the values obtained from recent measurement of the deceleration parameter today using the Pantheon supernova samples~\cite{Camarena:2019moy}. The interpretation of the inferred values of the deceleration parameters should be taken with caution because cosmography in the presence of structures at second order in redshift differs from the full expression by more than 100\% at about $z\sim 0.1$.

Finally, it is undeniable that the FLRW spacetime has played a big role in our current understanding of the Universe. Our results show 
that the tidal deformation of the observer spacetime region due to the gravitational interaction among nearby structures cannot be neglected. This interaction is characterised by tensors that vanish on the FLRW background spacetime. We have shown that not taking these into account when using the FLRW model alone to analyse cosmological observation leads to cosmological tensions.
We showed that the determination of Hubble rate from the analysis of the local distance ladder for the SNIa peak magnitudes takes into account the impact of the tidal deformation of the local spacetime consistently.

\section*{Acknowledgement}
I benefited immensely from discussions with Chris Clarkson, Robert Crittenden, Pierre Fleury,  Emir Gumrukcuoglu, Asta Heinesen, Mathew Hull, Antony Lewis, Kazuya Koyama and  Daniela Saadeh. The works by G.F.R. Ellis on this topic have been very influential. 
OU  is supported by the 
UK Science \& Technology Facilities Council (STFC) Consolidated Grants Grant ST/S000550/1 and the South African Square Kilometre Array Project. The perturbation theory computations in this paper  were done with the help of tensor algebra software xPand \cite{Pitrou:2013hga} which is based on xPert~\cite{Brizuela:2008ra}.
I made use of COLOSSUS (A python toolkit for cosmology, large-scale structure, and dark matter halos) developed by Benedikt Diemer for computations involving the dark matter halo profiles~\cite{Diemer:2017bwl}.

\section*{Data Access Statement}

{No new data were generated or analysed in support of this research.}

%\newpage
\appendix

\section{Covariant cosmography on the past lightcone}\label{sec:KristainandSachs}

The distance-redshift relation on arbitrary spacetime  may be expanded in Taylor series up to any order following the formalism developed by Kristain and Sachs ~\cite{1966ApJ...143..379K}.
The formalism uses the null focusing equation to propagate the deviation vector, $\xi^a$,  with the tail on the  central ray, $L$ and head on the nearby geodesics from a source with 4-velocity $u^a_s$ to the observer with 4-velocity $u^a_o$
\begin{equation}\label{eq:GeoDeqn}
\frac{\d^2\xi^a}{\d\lambda^2}=-R^a{}_{cbd}k^ck^d\xi^b\,\,,
\end{equation}
where  $R^a{}_{cbd}$   is the Riemann  curvature tensor, $\xi^a$ lives on the screen space and $k^a$ is the tangent vector to the null geodesics. 
%
%The vector $\d\xi^a/\d\lambda$ is orthogonal to $u^a$ and $k^a$. 
The solution to equation \eqref{eq:GeoDeqn} needs to satisfy the following initial conditions  at source 
%$x^a_1$ and $x^a_2$ (the positions $x^a_1$ and $x^a_2$ will later be replaced by `s' and `o' respectively),
\begin{eqnarray}\label{eq:InitalCond}
 (\xi^a)_s&=&0,
\qquad
\left (\frac{\d\xi^a}{\d\lambda}u_a\right)_s =0,
\qquad
\left(\frac{\d\xi^a}{\d\lambda}k_a\right)_s=0,
\qquad
\pi\left( \frac{\d\xi^a}{\d\lambda}\frac{\d\xi_a}{\d\lambda}\right)_s= - (u_a k^a)_s d\Omega\,.
\end{eqnarray}
Equation \eqref{eq:GeoDeqn} is valid provided there are no focal points between the observer and the source. 
 Expanding $(\xi_a\xi_b)$ in Taylor series in $\lambda$ up to fifth order, gives
\begin{eqnarray}
(\xi_a\xi_b)_o &=&\left(\xi_a \xi_b\right)_s + \left[\frac{\d}{\d\lambda}(\xi_a\xi_b)\right]_s\lambda_o+ \frac{1}{2}\left[\frac{\d^2}{\d\lambda^2}(\xi_a\xi_b)\right]_s\lambda_o^2
+  \mathcal{O} ( \lambda_o^3)\,.
\end{eqnarray}
Using  equation \eqref{eq:GeoDeqn} and the boundary conditions (equation \eqref{eq:InitalCond}), give
\begin{eqnarray}
(\xi_a\xi_b)_o &=&\lambda_o^2\left(\frac{\d\xi^c}{\d\lambda}\frac{\d\xi_c}{\d\lambda}\right)_s\left(g_{ab}-\frac{1}{3}k^ck^dR_{acbd}\lambda_o^2+ \mathcal{O}( \lambda_o^3)
\right)\,.
\end{eqnarray}
 The  area element is defined $dA=\sqrt{(\xi_a\xi_b)_o}d\Omega$, therefore
 \begin{eqnarray}\label{eq:Inftyarea}
 dA_o&=& d\Omega \lambda_o^2 (u_ak^a)^2_s \left(1-\frac{1}{6} k^ck^j R_{cj} \lambda_o^2
 + \mathcal{O}( \lambda_2^3)\right)\,.
  \end{eqnarray}
To make contact with observation  we expand the redshift in $\lambda$ 
\begin{eqnarray}\label{eq:Redshift}
\left(1+z\right)
=\frac{E_s}{E_o}&=&\frac{1}{(k^au_a)_0}\bigg[k^au_a+ k^ak^b\nabla_au_b \lambda + \frac{1}{2}k^ak^bk^c\nabla_a\nabla_b u_c \lambda^2+
 \mathcal{O}( \lambda^3)\bigg]\,.
\end{eqnarray}
  %
%Now using equation \eqref{eq:Inftyarea} and setting  $\lambda_2=\lambda_o$, $x^a_2=x^a_o=0$, $x^a_1=x^a_s$, $dA_2=dA_o$,  and 
Using  the  fact that the area distance  is the ratio of the cross-sectional area at the observer to the cross-sectional area  at the source according to 
  \begin{equation}\label{eq:areadistancedef}
  d^2_A=\frac{dA_0}{d\Omega}\frac{(u_ak^a)_o^2}{(u_ak^a)^2_s}\,.
  \end{equation}
Using  equation \eqref{eq:Inftyarea}  in equation \eqref{eq:areadistancedef} we find
  \begin{eqnarray}\label{eq:AngDAffcne}
  d^2_A&=&(u_ak^a)_0^2 \lambda_0^2 \left\{1-\frac{1}{6} k^ck^d R_{cd} \lambda_0^2+
    \mathcal{O}( \lambda_0^3)\right\}\,.
\end{eqnarray}
Inverting the series is equation \eqref{eq:AngDAffcne} gives  the affine parameter in terms of the area distance,
\begin{eqnarray}\label{eq:lambda0}
\lambda_0&=&d_A\left[1+\frac{1}{12}K^cK^dR_{cd} d^2_A +
  \mathcal{O}(d_A^3)\right]\,,
\end{eqnarray}
%\frac{D_A}{(u_ak^a)_o}
where we defined  normalised photon tangent vector  $K^a:=\frac{k^a}{(u_ak^a)_0}=-u^a+n^a$.
Then putting equation  \eqref{eq:lambda0} in equation \eqref{eq:Redshift} gives
 \begin{eqnarray}\label{eq:zDrelation}
z&=&\left.K^aK^b\nabla_au_b\right|_o d_A + \frac{1}{2}\left.\left(K^aK^bK^c\nabla_a\nabla_bu_c\right)\right|_o d^2_A 
+   \mathcal{O}(d_A^3)\,.
\end{eqnarray}
%where the renormalized null vector is given by $K^a=-(u^a+e^a)$.  I
Inverting equation \eqref{eq:zDrelation} gives the area distance in terms of the redshift
\begin{eqnarray}
d_A&=&\frac{z}{K^cK^j\nabla_cu_j|_0}\left\{1-\left[\frac{1}{2}\frac{K^cK^jK^k\nabla_k\nabla_ju_c}{\left(K^cK^j\nabla_cu_c\right)^2}\right]_0z  + \mathcal{O}(z^2)\right\}\,.
\end{eqnarray} 
Using the reciprocity theorem 
$d_L =d_A(1+z)^2
$, we find the luminosity distance to be
\begin{eqnarray} 
d_L&=&\frac{z}{K^cK^j\nabla_ju_c|_0}\left\{1+\frac{1}{2}\left[4-\frac{K^cK^jK^k\nabla_k\nabla_ju_c}{\left(K^cK^j\nabla_cu_c\right)^2}\right]_0z   + \mathcal{O}(z^2)\right\}\,.
\end{eqnarray}

%\item{ \tt{Hubble rate of intercept fitting}}
%\bibliographystyle{$HOME/Dropbox/UWC_papers/Effective_fnl/Biassecondorder/JHEP}%{JHEP}%
%\bibliographystyle{apsrev4-1}
%\bibliography{$HOME/Dropbox/UWC_papers/q-dipole/draft/cosmoref.bib}

\begin{thebibliography}{100}

\bibitem{Martinez:1998yp}
V.~J. Martinez, M.-J. Pons-Borderia, R.~A. Moyeed, and M.~J. Graham, {\it
  {Searching for the scale of homogeneity}},  {\em Mon. Not. Roy. Astron. Soc.}
  {\bf 298} (1998) 1212, [\href{http://arxiv.org/abs/astro-ph/9804073}{{\tt
  astro-ph/9804073}}].

\bibitem{Hogg:2004vw}
D.~W. Hogg, D.~J. Eisenstein, M.~R. Blanton, N.~A. Bahcall, J.~Brinkmann, J.~E.
  Gunn, and D.~P. Schneider, {\it {Cosmic homogeneity demonstrated with
  luminous red galaxies}},  {\em Astrophys. J.} {\bf 624} (2005) 54--58,
  [\href{http://arxiv.org/abs/astro-ph/0411197}{{\tt astro-ph/0411197}}].

\bibitem{Ntelis:2018ctq}
P.~Ntelis, A.~Ealet, S.~Escoffier, J.-C. Hamilton, A.~J. Hawken, J.-M. Le~Goff,
  J.~Rich, and A.~Tilquin, {\it {The scale of cosmic homogeneity as a standard
  ruler}},  {\em JCAP} {\bf 12} (2018) 014,
  [\href{http://arxiv.org/abs/1810.09362}{{\tt arXiv:1810.09362}}].

\bibitem{Boehringer:2021mix}
H.~Boehringer, G.~Chon, and J.~Truemper, {\it {The Cosmic Large-Scale Structure
  in X-rays (CLASSIX) Cluster Survey - III. The Perseus-Pisces supercluster and
  the Southern Great Wall as traced by X-ray luminous galaxy clusters}},  {\em
  Astron. Astrophys.} {\bf 651} (2021) A16,
  [\href{http://arxiv.org/abs/2105.14051}{{\tt arXiv:2105.14051}}].

\bibitem{Nastas:2011AandA...532L...6N}
A.~{Nastasi}, R.~{Fassbender}, H.~{B{\"o}hringer}, R.~{{\v{S}}uhada},
  P.~{Rosati}, D.~{Pierini}, M.~{Verdugo}, J.~S. {Santos}, A.~D. {Schwope},
  A.~{de Hoon}, J.~{Kohnert}, G.~{Lamer}, M.~{M{\"u}hlegger}, and
  H.~{Quintana}, {\it {Discovery of the X-ray selected galaxy cluster XMMU
  J0338.8+0021 at z = 1.49. Indications of a young system with a brightest
  galaxy in formation}},  {\em \aap} {\bf 532} (Aug., 2011) L6,
  [\href{http://arxiv.org/abs/1106.5784}{{\tt arXiv:1106.5784}}].

\bibitem{Chadayammuri:2021tcp}
U.~Chadayammuri, J.~ZuHone, P.~Nulsen, D.~Nagai, S.~Felix, F.~Andrade-Santos,
  L.~King, and H.~Russell, {\it {Constraining merging galaxy clusters with
  X-ray and lensing simulations and observations: the case of Abell 2146}},
  {\em Mon. Not. Roy. Astron. Soc.} {\bf 509} (2021), no.~1 1201--1216,
  [\href{http://arxiv.org/abs/2108.05296}{{\tt arXiv:2108.05296}}].

\bibitem{Ellis:1987zz}
G.~F.~R. Ellis and W.~Stoeger, {\it {The 'fitting problem' in cosmology}},
  {\em Class. Quant. Grav.} {\bf 4} (1987) 1697--1729.

\bibitem{EllSto87}
G.~F.~R. {Ellis} and W.~{Stoeger}, {\it {The 'fitting problem' in cosmology}},
  {\em Class. Quant. Grav.} {\bf 4} (Nov., 1987) 1697--1729.

\bibitem{Clarkson:2011zq}
C.~Clarkson, G.~Ellis, J.~Larena, and O.~Umeh, {\it {Does the growth of
  structure affect our dynamical models of the universe? The averaging,
  backreaction and fitting problems in cosmology}},  {\em Rept. Prog. Phys.}
  {\bf 74} (2011) 112901, [\href{http://arxiv.org/abs/1109.2314}{{\tt
  arXiv:1109.2314}}].

\bibitem{Larena:2012vn}
J.~Larena, {\it {The fitting problem in a lattice Universe}},  {\em Springer
  Proc. Phys.} {\bf 157} (2014) 385--392,
  [\href{http://arxiv.org/abs/1210.2161}{{\tt arXiv:1210.2161}}].

\bibitem{Maartens:2011yx}
R.~Maartens, {\it {Is the Universe homogeneous?}},  {\em Phil. Trans. Roy. Soc.
  Lond. A} {\bf 369} (2011) 5115--5137,
  [\href{http://arxiv.org/abs/1104.1300}{{\tt arXiv:1104.1300}}].

\bibitem{Clarkson:2012bg}
C.~Clarkson, {\it {Establishing homogeneity of the universe in the shadow of
  dark energy}},  {\em Comptes Rendus Physique} {\bf 13} (2012) 682--718,
  [\href{http://arxiv.org/abs/1204.5505}{{\tt arXiv:1204.5505}}].

\bibitem{Riess:2016jrr}
A.~G. Riess et~al., {\it {A 2.4\% Determination of the Local Value of the
  Hubble Constant}},  {\em Astrophys. J.} {\bf 826} (2016), no.~1 56,
  [\href{http://arxiv.org/abs/1604.01424}{{\tt arXiv:1604.01424}}].

\bibitem{2019Natur.567..200P}
G.~{Pietrzy{\'n}ski}, D.~{Graczyk}, A.~{Gallenne}, W.~{Gieren}, I.~B.
  {Thompson}, B.~{Pilecki}, P.~{Karczmarek}, M.~{G{\'o}rski}, K.~{Suchomska},
  M.~{Taormina}, B.~{Zgirski}, P.~{Wielg{\'o}rski}, Z.~{Ko{\l}aczkowski},
  P.~{Konorski}, S.~{Villanova}, N.~{Nardetto}, P.~{Kervella}, F.~{Bresolin},
  R.~P. {Kudritzki}, J.~{Storm}, R.~{Smolec}, and W.~{Narloch}, {\it {A
  distance to the Large Magellanic Cloud that is precise to one per cent}},
  {\em \nat} {\bf 567} (Mar., 2019) 200--203,
  [\href{http://arxiv.org/abs/1903.08096}{{\tt arXiv:1903.08096}}].

\bibitem{Reid:2019tiq}
M.~J. Reid, D.~W. Pesce, and A.~G. Riess, {\it {An Improved Distance to NGC
  4258 and its Implications for the Hubble Constant}},  {\em Astrophys. J.
  Lett.} {\bf 886} (2019), no.~2 L27,
  [\href{http://arxiv.org/abs/1908.05625}{{\tt arXiv:1908.05625}}].

\bibitem{Freedman:2019jwv}
W.~L. {Freedman}, B.~F. {Madore}, D.~{Hatt}, T.~J. {Hoyt}, I.~S. {Jang}, R.~L.
  {Beaton}, C.~R. {Burns}, M.~G. {Lee}, A.~J. {Monson}, J.~R. {Neeley}, M.~M.
  {Phillips}, J.~A. {Rich}, and M.~{Seibert}, {\it {The Carnegie-Chicago Hubble
  Program. VIII. An Independent Determination of the Hubble Constant Based on
  the Tip of the Red Giant Branch}},  {\em \apj} {\bf 882} (Sept., 2019) 34,
  [\href{http://arxiv.org/abs/1907.05922}{{\tt arXiv:1907.05922}}].

\bibitem{Anand:2021sum}
G.~S. Anand, R.~B. Tully, L.~Rizzi, A.~G. Riess, and W.~Yuan, {\it {Comparing
  Tip of the Red Giant Branch Distance Scales: An Independent Reduction of the
  Carnegie-Chicago Hubble Program and the Value of the Hubble Constant}},
  \href{http://arxiv.org/abs/2108.00007}{{\tt arXiv:2108.00007}}.

\bibitem{Alcock:1979mp}
C.~Alcock and B.~Paczynski, {\it {An evolution free test for non-zero
  cosmological constant}},  {\em Nature} {\bf 281} (1979) 358--359.

\bibitem{Anderson:2013oza}
L.~Anderson et~al., {\it {The clustering of galaxies in the SDSS-III Baryon
  Oscillation Spectroscopic Survey: measuring $D_A$ and H at z = 0.57 from the
  baryon acoustic peak in the Data Release 9 spectroscopic Galaxy sample}},
  {\em Mon. Not. Roy. Astron. Soc.} {\bf 439} (2014), no.~1 83--101,
  [\href{http://arxiv.org/abs/1303.4666}{{\tt arXiv:1303.4666}}].

\bibitem{BOSS:2016wmc}
{\bf BOSS} Collaboration, S.~Alam et~al., {\it {The clustering of galaxies in
  the completed SDSS-III Baryon Oscillation Spectroscopic Survey: cosmological
  analysis of the DR12 galaxy sample}},  {\em Mon. Not. Roy. Astron. Soc.} {\bf
  470} (2017), no.~3 2617--2652, [\href{http://arxiv.org/abs/1607.03155}{{\tt
  arXiv:1607.03155}}].

\bibitem{Philcox:2020vvt}
O.~H.~E. Philcox, M.~M. Ivanov, M.~Simonovi\'c, and M.~Zaldarriaga, {\it
  {Combining Full-Shape and BAO Analyses of Galaxy Power Spectra: A
  1.6\textbackslash{}\% CMB-independent constraint on H$_0$}},  {\em JCAP} {\bf
  05} (2020) 032, [\href{http://arxiv.org/abs/2002.04035}{{\tt
  arXiv:2002.04035}}].

\bibitem{DiValentino:2021izs}
E.~Di~Valentino, O.~Mena, S.~Pan, L.~Visinelli, W.~Yang, A.~Melchiorri, D.~F.
  Mota, A.~G. Riess, and J.~Silk, {\it {In the realm of the Hubble
  tension\textemdash{}a review of solutions}},  {\em Class. Quant. Grav.} {\bf
  38} (2021), no.~15 153001, [\href{http://arxiv.org/abs/2103.01183}{{\tt
  arXiv:2103.01183}}].

\bibitem{Aghanim:2018eyx}
{\bf Planck} Collaboration, N.~Aghanim et~al., {\it {Planck 2018 results. VI.
  Cosmological parameters}},  {\em Astron. Astrophys.} {\bf 641} (2020) A6,
  [\href{http://arxiv.org/abs/1807.06209}{{\tt arXiv:1807.06209}}].

\bibitem{DES:2021jns}
{\bf DES} Collaboration, A.~C. Rosell et~al., {\it {Dark Energy Survey Year 3
  Results: Galaxy Sample for BAO Measurement}},
  \href{http://arxiv.org/abs/2107.05477}{{\tt arXiv:2107.05477}}.

\bibitem{1966ApJ...143..379K}
J.~{Kristian} and R.~K. {Sachs}, {\it {Observations in Cosmology}},  {\em \apj}
  {\bf 143} (Feb., 1966) 379--+.

\bibitem{Ellis:1966ta}
G.~F.~R. Ellis, {\it {Dynamics of pressure free matter in general relativity}},
   {\em J. Math. Phys.} {\bf 8} (1967) 1171--1194.

\bibitem{Ellis:1998ct}
G.~F.~R. Ellis and H.~van Elst, {\it {Cosmological models}},  {\em NATO Adv.
  Study Inst. Ser. C. Math. Phys. Sci.} {\bf 541} (1999) 1--116,
  [\href{http://arxiv.org/abs/gr-qc/9812046}{{\tt gr-qc/9812046}}].

\bibitem{Pirani2}
F.~A.~E. {Pirani}, {\em {INTRODUCTION TO GRAVITATIONAL RADIATION THEORY (Notes
  by J. J. J. Marek and the Lecturer)}}, pp.~249--+.
\newblock 1965.

\bibitem{Thorne1980}
K.~S. {Thorne}, {\it {Multipole expansions of gravitational radiation}},  {\em
  Reviews of Modern Physics} {\bf 52} (Apr., 1980) 299--340.

\bibitem{Witten:2019qhl}
E.~Witten, {\it {Light Rays, Singularities, and All That}},  {\em Rev. Mod.
  Phys.} {\bf 92} (2020), no.~4 045004,
  [\href{http://arxiv.org/abs/1901.03928}{{\tt arXiv:1901.03928}}].

\bibitem{Li:2007eg}
Y.-S. Li and S.~D.~M. White, {\it {Masses for the Local Group and the Milky
  Way}},  {\em Mon. Not. Roy. Astron. Soc.} {\bf 384} (2008) 1459--1468,
  [\href{http://arxiv.org/abs/0710.3740}{{\tt arXiv:0710.3740}}].

\bibitem{Carlesi:2016eud}
E.~Carlesi, Y.~Hoffman, J.~G. Sorce, and S.~Gottl\"ober, {\it {Constraining the
  mass of the Local Group}},  {\em Mon. Not. Roy. Astron. Soc.} {\bf 465}
  (2017), no.~4 4886--4894, [\href{http://arxiv.org/abs/1611.08078}{{\tt
  arXiv:1611.08078}}].

\bibitem{Li:2021sqb}
Z.-Z. Li and J.~Han, {\it {The Outermost Edges of the Milky Way Halo from
  Galaxy Kinematics}},  {\em Astrophys. J. Lett.} {\bf 915} (2021), no.~1 L18,
  [\href{http://arxiv.org/abs/2105.04978}{{\tt arXiv:2105.04978}}].

\bibitem{Efstathiou:2021ocp}
G.~Efstathiou, {\it {To H0 or not to H0?}},  {\em Mon. Not. Roy. Astron. Soc.}
  {\bf 505} (2021), no.~3 3866--3872,
  [\href{http://arxiv.org/abs/2103.08723}{{\tt arXiv:2103.08723}}].

\bibitem{Camarena:2019moy}
D.~Camarena and V.~Marra, {\it {Local determination of the Hubble constant and
  the deceleration parameter}},  {\em Phys. Rev. Res.} {\bf 2} (2020), no.~1
  013028, [\href{http://arxiv.org/abs/1906.11814}{{\tt arXiv:1906.11814}}].

\bibitem{Umeh:2022hab}
O.~Umeh, {\it {The art of building a smooth cosmic distance ladder in a
  perturbed universe}},  \href{http://arxiv.org/abs/2201.11089}{{\tt
  arXiv:2201.11089}}.

\bibitem{Riess:2021jrx}
A.~G. Riess et~al., {\it {A Comprehensive Measurement of the Local Value of the
  Hubble Constant with 1 km/s/Mpc Uncertainty from the Hubble Space Telescope
  and the SH0ES Team}},  \href{http://arxiv.org/abs/2112.04510}{{\tt
  arXiv:2112.04510}}.

\bibitem{2010fimv.book..267S}
D.~J. {Schwarz}, {\em {Thoughts on the Cosmological Principle}}, pp.~267--276.
\newblock 2010.

\bibitem{Sugiyama:2021axw}
S.~Sugiyama et~al., {\it {HSC Year 1 cosmology results with the minimal bias
  method: HSC$\times$BOSS galaxy-galaxy weak lensing and BOSS galaxy
  clustering}},  \href{http://arxiv.org/abs/2111.10966}{{\tt
  arXiv:2111.10966}}.

\bibitem{Zhao:2021ahg}
C.~Zhao et~al., {\it {The completed SDSS-IV extended Baryon Oscillation
  Spectroscopic Survey: Cosmological implications from multi-tracer BAO
  analysis with galaxies and voids}},
  \href{http://arxiv.org/abs/2110.03824}{{\tt arXiv:2110.03824}}.

\bibitem{McClure:2007hy}
M.~L. McClure and C.~Hellaby, {\it {The Metric of the Cosmos II: Accuracy,
  Stability, and Consistency}},  {\em Phys. Rev. D} {\bf 78} (2008) 044005,
  [\href{http://arxiv.org/abs/0709.0875}{{\tt arXiv:0709.0875}}].

\bibitem{Hellaby:2008pp}
C.~Hellaby and A.~H.~A. Alfedeel, {\it {Solving the Observer Metric}},  {\em
  Phys. Rev. D} {\bf 79} (2009) 043501,
  [\href{http://arxiv.org/abs/0811.1676}{{\tt arXiv:0811.1676}}].

\bibitem{Bolejko:2011ys}
K.~Bolejko, C.~Hellaby, and A.~H.~A. Alfedeel, {\it {The Metric of the Cosmos
  from Luminosity and Age Data}},  {\em JCAP} {\bf 09} (2011) 011,
  [\href{http://arxiv.org/abs/1102.3370}{{\tt arXiv:1102.3370}}].

\bibitem{Lu:2007gr}
T.~H.-C. Lu and C.~Hellaby, {\it {Obtaining the spacetime metric from
  cosmological observations}},  {\em Class. Quant. Grav.} {\bf 24} (2007)
  4107--4132, [\href{http://arxiv.org/abs/0705.1060}{{\tt arXiv:0705.1060}}].

\bibitem{Ellis:1984bqf}
G.~F.~R. Ellis, {\it {Relativistic Cosmology: Its Nature, Aims and Problems}},
  {\em Fundam. Theor. Phys.} {\bf 9} (1984) 215--288.

\bibitem{Perlick:2010zh}
V.~Perlick, {\it {Gravitational Lensing from a Spacetime Perspective}},  {\em
  Living Rev.Rel.} (2010) [\href{http://arxiv.org/abs/1010.3416}{{\tt
  arXiv:1010.3416}}].

\bibitem{Rizzi:2007ni}
L.~Rizzi, R.~B. Tully, D.~Makarov, L.~Makarova, A.~E. Dolphin, S.~Sakai, and
  E.~J. Shaya, {\it {Tip of the Red Giant Branch Distances. 2. Zero-Point
  Calibration}},  {\em Astrophys. J.} {\bf 661} (2007) 815--829,
  [\href{http://arxiv.org/abs/astro-ph/0701518}{{\tt astro-ph/0701518}}].

\bibitem{2015PASP..127..102M}
A.~W. {Mann} and K.~{von Braun}, {\it {Revised Filter Profiles and Zero Points
  for Broadband Photometry}},  {\em \pasp} {\bf 127} (Feb., 2015) 102,
  [\href{http://arxiv.org/abs/1412.1474}{{\tt arXiv:1412.1474}}].

\bibitem{Madore:2021ktx}
B.~F. Madore, W.~L. Freedman, and A.~Lee, {\it {Astrophysical Distance Scale
  IV. Preliminary Zero-Point Calibration of the JAGB Method in the HST/WFC3-IR
  Broad J-Band (F110W) Filter}},  \href{http://arxiv.org/abs/2112.06968}{{\tt
  arXiv:2112.06968}}.

\bibitem{Heinesen:2020bej}
A.~Heinesen, {\it {Multipole decomposition of the general luminosity distance
  'Hubble law' -- a new framework for observational cosmology}},
  \href{http://arxiv.org/abs/2010.06534}{{\tt arXiv:2010.06534}}.

\bibitem{Etheringto:1933}
I.~Etherington, {\it Lx. on the definition of distance in general relativity},
  {\em The London, Edinburgh, and Dublin Philosophical Magazine and Journal of
  Science} {\bf 15} (1933), no.~100 761--773,
  [\href{http://arxiv.org/abs/https://doi.org/10.1080/14786443309462220}{{\tt
  https://doi.org/10.1080/14786443309462220}}].

\bibitem{Ellis1971grc..conf..104E}
G.~F.~R. {Ellis}, {\it {Relativistic cosmology.}},  in {\em General Relativity
  and Cosmology} (R.~K. {Sachs}, ed.), pp.~104--182, 1971.

\bibitem{Ellis2009}
G.~F.~R. Ellis, {\it Republication of: Relativistic cosmology},  {\em General
  Relativity and Gravitation} {\bf 41} (Mar, 2009) 581--660.

\bibitem{Peterson:2021hel}
E.~R. Peterson et~al., {\it {The Pantheon+ Analysis: Evaluating Peculiar
  Velocity Corrections in Cosmological Analyses with Nearby Type Ia
  Supernovae}},  \href{http://arxiv.org/abs/2110.03487}{{\tt
  arXiv:2110.03487}}.

\bibitem{Sussman:2017otk}
R.~A. Sussman, J.~C. Hidalgo, I.~Delgado~Gaspar, and G.~Germ\'an, {\it
  {Nonspherical Szekeres models in the language of cosmological
  perturbations}},  {\em Phys. Rev. D} {\bf 95} (2017), no.~6 064033,
  [\href{http://arxiv.org/abs/1701.00819}{{\tt arXiv:1701.00819}}].

\bibitem{Macpherson:2021gbh}
H.~J. Macpherson and A.~Heinesen, {\it {Luminosity distance and anisotropic
  sky-sampling at low redshifts: A numerical relativity study}},  {\em Phys.
  Rev. D} {\bf 104} (2021) 023525, [\href{http://arxiv.org/abs/2103.11918}{{\tt
  arXiv:2103.11918}}]. [Erratum: Phys.Rev.D 104, 109901 (2021)].

\bibitem{EGS1968}
J.~{Ehlers}, P.~{Geren}, and R.~K. {Sachs}, {\it {Isotropic solutions of the
  Einstein-Liouville equations.}},  {\em Journal of Mathematical Physics} {\bf
  9} (1968) 1344--1349.

\bibitem{Ehlers:1993gf}
J.~Ehlers, {\it {Contributions to the relativistic mechanics of continuous
  media}},  {\em Gen.Rel.Grav.} {\bf 25} (1993) 1225--1266.

\bibitem{Ellis:1990gi}
G.~F.~R. Ellis, M.~Bruni, and J.~Hwang, {\it {Density gradient-vorticity
  relation in perfect fluid Robertson-Walker perturbations}},  {\em Phys. Rev.}
  {\bf D42} (1990) 1035--1046.

\bibitem{1981MNRAS.194..439T}
K.~S. {Thorne}, {\it {Relativistic radiative transfer - Moment formalisms}},
  {\em \mnras} {\bf 194} (Feb., 1981) 439--473.

\bibitem{2000AnPhy.282..285G}
T.~{Gebbie} and G.~F.~R. {Ellis}, {\it {1+3 covariant cosmic microwave
  background anisotropies. I. Algebraic relations for mode and multipole
  expansions.}},  {\em Annals of Physics} {\bf 282} (June, 2000) 285--320,
  [\href{http://arxiv.org/abs/astro-ph/9}{{\tt astro-ph/9}}].

\bibitem{Ellis1983455}
G.~F.~R. Ellis, D.~R. Matravers, and R.~Treciokas, {\it Anisotropic solutions
  of the einstein-boltzmann equations: I. general formalism},  {\em Annals of
  Physics} {\bf 150} (1983), no.~2 455 -- 486.

\bibitem{Gebbie:1998fe}
T.~Gebbie and G.~F.~R. Ellis, {\it {GIC approach to cosmic background radiation
  anisotropies. Part 1.}},  {\em Annals Phys.} {\bf 282} (2006) 285--320,
  [\href{http://arxiv.org/abs/astro-ph/9804316}{{\tt astro-ph/9804316}}].

\bibitem{Spencer1970}
A.~J.~M. Spencer, {\it A note on the decomposition of tensors into traceless
  symmetric tensors},  {\em International Journal of Engineering Science} {\bf
  8} (1970), no.~6 475 -- 481.

\bibitem{Thorne:1980ru}
K.~Thorne, {\it {Multipole Expansions of Gravitational Radiation}},  {\em
  Rev.Mod.Phys.} {\bf 52} (1980) 299--339.

\bibitem{Liotta2000L}
S.~{Liotta}, {\it {Moment equations for electrons in semiconductors: comparison
  of spherical harmonics and full moments}},  {\em Solid State Electronics}
  {\bf 44} (Jan., 2000) 95--103.

\bibitem{Jaric2003}
J.~P. Jaric, {\it On the decomposition of symmetric tensors into traceless
  symmetric tensors},  {\em International Journal of Engineering Science} {\bf
  41} (2003), no.~18 2123 -- 2141.

\bibitem{Clarkson:2011uk}
C.~Clarkson and O.~Umeh, {\it {Is backreaction really small within concordance
  cosmology?}},  {\em Class. Quant. Grav.} {\bf 28} (2011) 164010,
  [\href{http://arxiv.org/abs/1105.1886}{{\tt arXiv:1105.1886}}].

\bibitem{Umeh:2013UCT}
O.~Umeh, {\em "The influence of structure formation on the evolution of the
  universe."}.
\newblock PhD thesis, University of Cape Town, Faculty of Science, Department
  of Mathematics and Applied Mathematics,
  https://open.uct.ac.za/handle/11427/4938, 2013.

\bibitem{Nadathur:2019mct}
S.~Nadathur, P.~M. Carter, W.~J. Percival, H.~A. Winther, and J.~Bautista, {\it
  {Beyond BAO: Improving cosmological constraints from BOSS data with
  measurement of the void-galaxy cross-correlation}},  {\em Phys. Rev. D} {\bf
  100} (2019), no.~2 023504, [\href{http://arxiv.org/abs/1904.01030}{{\tt
  arXiv:1904.01030}}].

\bibitem{Beutler:2014yhv}
{\bf BOSS} Collaboration, F.~Beutler et~al., {\it {The clustering of galaxies
  in the SDSS-III Baryon Oscillation Spectroscopic Survey: signs of neutrino
  mass in current cosmological data sets}},  {\em Mon. Not. Roy. Astron. Soc.}
  {\bf 444} (2014), no.~4 3501--3516,
  [\href{http://arxiv.org/abs/1403.4599}{{\tt arXiv:1403.4599}}].

\bibitem{Umeh:2014ana}
O.~Umeh, C.~Clarkson, and R.~Maartens, {\it {Nonlinear relativistic corrections
  to cosmological distances, redshift and gravitational lensing magnification.
  II - Derivation}},  {\em Class. Quant. Grav.} {\bf 31} (2014) 205001,
  [\href{http://arxiv.org/abs/1402.1933}{{\tt arXiv:1402.1933}}].

\bibitem{2012MNRAS.425.2049H}
Y.~{Hoffman}, O.~{Metuki}, G.~{Yepes}, S.~{Gottl{\"o}ber}, J.~E.
  {Forero-Romero}, N.~I. {Libeskind}, and A.~{Knebe}, {\it {A kinematic
  classification of the cosmic web}},  {\em \mnras} {\bf 425} (Sept., 2012)
  2049--2057, [\href{http://arxiv.org/abs/1201.3367}{{\tt arXiv:1201.3367}}].

\bibitem{Forero-Romero:2014fna}
J.~E. Forero-Romero, S.~Contreras, and N.~Padilla, {\it {Cosmic web alignments
  with the shape, angular momentum and peculiar velocities of dark matter
  haloes}},  {\em Mon. Not. Roy. Astron. Soc.} {\bf 443} (2014), no.~2
  1090--1102, [\href{http://arxiv.org/abs/1406.0508}{{\tt arXiv:1406.0508}}].

\bibitem{MacEllis70}
M.~A.~H. {MacCallum} and G.~F.~R. {Ellis}, {\it {A class of homogeneous
  cosmological models: II. Observations}},  {\em Communications in Mathematical
  Physics} {\bf 19} (Mar., 1970) 31--64.

\bibitem{Freedman:2021ahq}
W.~L. Freedman, {\it {Measurements of the Hubble Constant: Tensions in
  Perspective}},  {\em Astrophys. J.} {\bf 919} (2021), no.~1 16,
  [\href{http://arxiv.org/abs/2106.15656}{{\tt arXiv:2106.15656}}].

\bibitem{Freedman:2020dne}
W.~L. Freedman, B.~F. Madore, T.~Hoyt, I.~S. Jang, R.~Beaton, M.~G. Lee,
  A.~Monson, J.~Neeley, and J.~Rich, {\it {Calibration of the Tip of the Red
  Giant Branch (TRGB)}},  \href{http://arxiv.org/abs/2002.01550}{{\tt
  arXiv:2002.01550}}.

\bibitem{Riess:2018byc}
A.~G. Riess et~al., {\it {Milky Way Cepheid Standards for Measuring Cosmic
  Distances and Application to Gaia DR2: Implications for the Hubble
  Constant}},  {\em Astrophys. J.} {\bf 861} (2018), no.~2 126,
  [\href{http://arxiv.org/abs/1804.10655}{{\tt arXiv:1804.10655}}].

\bibitem{2018ApJ...855..136R}
A.~G. {Riess}, S.~{Casertano}, W.~{Yuan}, L.~{Macri}, J.~{Anderson}, J.~W.
  {MacKenty}, J.~B. {Bowers}, K.~I. {Clubb}, A.~V. {Filippenko}, D.~O. {Jones},
  and B.~E. {Tucker}, {\it {New Parallaxes of Galactic Cepheids from Spatially
  Scanning the Hubble Space Telescope: Implications for the Hubble Constant}},
  {\em \apj} {\bf 855} (Mar., 2018) 136,
  [\href{http://arxiv.org/abs/1801.01120}{{\tt arXiv:1801.01120}}].

\bibitem{Riess:2006fw}
A.~G. Riess et~al., {\it {New Hubble Space Telescope Discoveries of Type Ia
  Supernovae at z\ensuremath{>}=1: Narrowing Constraints on the Early Behavior
  of Dark Energy}},  {\em Astrophys. J.} {\bf 659} (2007) 98--121,
  [\href{http://arxiv.org/abs/astro-ph/0611572}{{\tt astro-ph/0611572}}].

\bibitem{Riess:2019cxk}
A.~G. Riess, S.~Casertano, W.~Yuan, L.~M. Macri, and D.~Scolnic, {\it {Large
  Magellanic Cloud Cepheid Standards Provide a 1\% Foundation for the
  Determination of the Hubble Constant and Stronger Evidence for Physics beyond
  $\Lambda$CDM}},  {\em Astrophys. J.} {\bf 876} (2019), no.~1 85,
  [\href{http://arxiv.org/abs/1903.07603}{{\tt arXiv:1903.07603}}].

\bibitem{Riess:2020fzl}
A.~G. Riess, S.~Casertano, W.~Yuan, J.~B. Bowers, L.~Macri, J.~C. Zinn, and
  D.~Scolnic, {\it {Cosmic Distances Calibrated to 1\% Precision with Gaia EDR3
  Parallaxes and Hubble Space Telescope Photometry of 75 Milky Way Cepheids
  Confirm Tension with $\Lambda$CDM}},  {\em Astrophys. J. Lett.} {\bf 908}
  (2021), no.~1 L6, [\href{http://arxiv.org/abs/2012.08534}{{\tt
  arXiv:2012.08534}}].

\bibitem{Aubourg:2014yra}
E.~Aubourg et~al., {\it {Cosmological implications of baryon acoustic
  oscillation measurements}},  {\em Phys. Rev. D} {\bf 92} (2015), no.~12
  123516, [\href{http://arxiv.org/abs/1411.1074}{{\tt arXiv:1411.1074}}].

\bibitem{Camarena:2019rmj}
D.~Camarena and V.~Marra, {\it {A new method to build the (inverse) distance
  ladder}},  {\em Mon. Not. Roy. Astron. Soc.} {\bf 495} (2020), no.~3
  2630--2644, [\href{http://arxiv.org/abs/1910.14125}{{\tt arXiv:1910.14125}}].

\bibitem{Umeh:2010pr}
O.~Umeh, J.~Larena, and C.~Clarkson, {\it {The Hubble rate in averaged
  cosmology}},  {\em JCAP} {\bf 1103} (2011) 029,
  [\href{http://arxiv.org/abs/1011.3959}{{\tt arXiv:1011.3959}}].

\bibitem{Ellis:2010fr}
G.~F.~R. Ellis and W.~R. Stoeger, {\it {The Evolution of Our Local Cosmic
  Domain: Effective Causal Limits}},  {\em Mon. Not. Roy. Astron. Soc.} {\bf
  398} (2009) 1527--1536, [\href{http://arxiv.org/abs/1001.4572}{{\tt
  arXiv:1001.4572}}].

\bibitem{Ellis:2020kry}
G.~F.~R. Ellis, {\it {The Causal Closure of Physics in Real World Contexts}},
  {\em Found. Phys.} {\bf 50} (2020), no.~10 1057--1097,
  [\href{http://arxiv.org/abs/2006.00972}{{\tt arXiv:2006.00972}}].

\bibitem{Karachentsev:2008st}
I.~D. Karachentsev, O.~G. Kashibadze, D.~I. Makarov, and R.~B. Tully, {\it {The
  Hubble flow around the Local Group}},  {\em Mon. Not. Roy. Astron. Soc.} {\bf
  393} (2009) 1265, [\href{http://arxiv.org/abs/0811.4610}{{\tt
  arXiv:0811.4610}}].

\bibitem{1989AandA...223...89E}
J.~{Einasto} and U.~{Haud}, {\it {Galactic models with massive corona. I -
  Method. II - Galaxy}},  {\em \aap} {\bf 223} (Oct., 1989) 89--106.

\bibitem{Navarro:1995iw}
J.~F. Navarro, C.~S. Frenk, and S.~D.~M. White, {\it {The Structure of cold
  dark matter halos}},  {\em Astrophys. J.} {\bf 462} (1996) 563--575,
  [\href{http://arxiv.org/abs/astro-ph/9508025}{{\tt astro-ph/9508025}}].

\bibitem{Diemer:2017bwl}
B.~Diemer, {\it {COLOSSUS: A python toolkit for cosmology, large-scale
  structure, and dark matter halos}},  {\em Astrophys. J. Suppl.} {\bf 239}
  (2018), no.~2 35, [\href{http://arxiv.org/abs/1712.04512}{{\tt
  arXiv:1712.04512}}].

\bibitem{More:2015ufa}
S.~More, B.~Diemer, and A.~Kravtsov, {\it {The splashback radius as a physical
  halo boundary and the growth of halo mass}},  {\em Astrophys. J.} {\bf 810}
  (2015), no.~1 36, [\href{http://arxiv.org/abs/1504.05591}{{\tt
  arXiv:1504.05591}}].

\bibitem{Umeh:2021xqm}
O.~Umeh, R.~Maartens, H.~Padmanabhan, and S.~Camera, {\it {The effect of finite
  halo size on the clustering of neutral hydrogen}},  {\em JCAP} {\bf 06}
  (2021) 027, [\href{http://arxiv.org/abs/2102.06116}{{\tt arXiv:2102.06116}}].

\bibitem{Kazantzidis:2005su}
S.~Kazantzidis, A.~R. Zentner, and A.~V. Kravtsov, {\it {The robustness of dark
  matter density profiles in dissipationless mergers}},  {\em Astrophys. J.}
  {\bf 641} (2006) 647--664, [\href{http://arxiv.org/abs/astro-ph/0510583}{{\tt
  astro-ph/0510583}}].

\bibitem{Valluri:2006vi}
M.~Valluri, I.~M. Vass, S.~Kazantzidis, A.~V. Kravtsov, and C.~L. Bohn, {\it
  {On relaxation processes in collisionless mergers}},  {\em Astrophys. J.}
  {\bf 658} (2007) 731, [\href{http://arxiv.org/abs/astro-ph/0609612}{{\tt
  astro-ph/0609612}}].

\bibitem{Carucci:2014ema}
I.~P. Carucci, M.~Sparre, S.~H. Hansen, and M.~Joyce, {\it {Particle ejection
  during mergers of dark matter halos}},  {\em JCAP} {\bf 06} (2014) 057,
  [\href{http://arxiv.org/abs/1405.6725}{{\tt arXiv:1405.6725}}].

\bibitem{Diemer:2014xya}
B.~Diemer and A.~V. Kravtsov, {\it {Dependence of the outer density profiles of
  halos on their mass accretion rate}},  {\em Astrophys. J.} {\bf 789} (2014)
  1, [\href{http://arxiv.org/abs/1401.1216}{{\tt arXiv:1401.1216}}].

\bibitem{1999AandARv...9..273V}
S.~{van den Bergh}, {\it {The local group of galaxies}},  {\em \aapr} {\bf 9}
  (Jan., 1999) 273--318.

\bibitem{1930AnHar..85..143L}
H.~S. {Leavitt} and H.~{Shapley}, {\it {Standards of magnitude for the
  astrographic catalogue}},  {\em Annals of Harvard College Observatory} {\bf
  85} (Jan., 1930) 143--156.

\bibitem{Camarena:2021jlr}
D.~Camarena and V.~Marra, {\it {On the use of the local prior on the absolute
  magnitude of Type Ia supernovae in cosmological inference}},  {\em Mon. Not.
  Roy. Astron. Soc.} {\bf 504} (2021) 5164--5171,
  [\href{http://arxiv.org/abs/2101.08641}{{\tt arXiv:2101.08641}}].

\bibitem{Dainotti:2021pqg}
M.~G. Dainotti, B.~De~Simone, T.~Schiavone, G.~Montani, E.~Rinaldi, and
  G.~Lambiase, {\it {On the Hubble constant tension in the SNe Ia Pantheon
  sample}},  {\em Astrophys. J.} {\bf 912} (2021), no.~2 150,
  [\href{http://arxiv.org/abs/2103.02117}{{\tt arXiv:2103.02117}}].

\bibitem{Niedermann:2021vgd}
F.~Niedermann and M.~S. Sloth, {\it {Hot new early dark energy}},  {\em Phys.
  Rev. D} {\bf 105} (2022), no.~6 063509,
  [\href{http://arxiv.org/abs/2112.00770}{{\tt arXiv:2112.00770}}].

\bibitem{Knox:2019rjx}
L.~Knox and M.~Millea, {\it {Hubble constant hunter\textquoteright{}s guide}},
  {\em Phys. Rev. D} {\bf 101} (2020), no.~4 043533,
  [\href{http://arxiv.org/abs/1908.03663}{{\tt arXiv:1908.03663}}].

\bibitem{Abdalla:2022yfr}
E.~Abdalla et~al., {\it {Cosmology intertwined: A review of the particle
  physics, astrophysics, and cosmology associated with the cosmological
  tensions and anomalies}},  {\em JHEAp} {\bf 34} (2022) 49--211,
  [\href{http://arxiv.org/abs/2203.06142}{{\tt arXiv:2203.06142}}].

\bibitem{Adler:2019fnp}
S.~L. Adler, {\it {Implications of a frame dependent dark energy for the
  spacetime metric, cosmography, and effective Hubble constant}},  {\em Phys.
  Rev. D} {\bf 100} (2019), no.~12 123503,
  [\href{http://arxiv.org/abs/1905.08228}{{\tt arXiv:1905.08228}}].

\bibitem{Capozziello:2020nyq}
S.~Capozziello, M.~Benetti, and A.~D. A.~M. Spallicci, {\it {Addressing the
  cosmological $H_0$ tension by the Heisenberg uncertainty}},  {\em Found.
  Phys.} {\bf 50} (2020), no.~9 893--899,
  [\href{http://arxiv.org/abs/2007.00462}{{\tt arXiv:2007.00462}}].

\bibitem{Benevento:2022cql}
G.~Benevento, J.~A. Kable, G.~E. Addison, and C.~L. Bennett, {\it {An
  exploration of an early gravity transition in light of cosmological
  tensions}},  \href{http://arxiv.org/abs/2202.09356}{{\tt arXiv:2202.09356}}.

\bibitem{Zumalacarregui:2020cjh}
M.~Zumalacarregui, {\it {Gravity in the Era of Equality: Towards solutions to
  the Hubble problem without fine-tuned initial conditions}},  {\em Phys. Rev.
  D} {\bf 102} (2020), no.~2 023523,
  [\href{http://arxiv.org/abs/2003.06396}{{\tt arXiv:2003.06396}}].

\bibitem{Dainotti:2022bzg}
M.~G. Dainotti, B.~De~Simone, T.~Schiavone, G.~Montani, E.~Rinaldi,
  G.~Lambiase, M.~Bogdan, and S.~Ugale, {\it {On the Evolution of the Hubble
  Constant with the SNe Ia Pantheon Sample and Baryon Acoustic Oscillations: A
  Feasibility Study for GRB-Cosmology in 2030}},  {\em Galaxies} {\bf 10}
  (2022), no.~1 24, [\href{http://arxiv.org/abs/2201.09848}{{\tt
  arXiv:2201.09848}}].

\bibitem{Pichon:1999tk}
C.~Pichon and F.~Bernardeau, {\it {Vorticity generation in large scale
  structure caustics}},  {\em Astron. Astrophys.} {\bf 343} (1999) 663,
  [\href{http://arxiv.org/abs/astro-ph/9902142}{{\tt astro-ph/9902142}}].

\bibitem{Lombriser:2019ahl}
L.~Lombriser, {\it {Consistency of the local Hubble constant with the cosmic
  microwave background}},  {\em Phys. Lett. B} {\bf 803} (2020) 135303,
  [\href{http://arxiv.org/abs/1906.12347}{{\tt arXiv:1906.12347}}].

\bibitem{Kasai:2019yqn}
M.~Kasai and T.~Futamase, {\it {A possible solution to the Hubble constant
  discrepancy -- Cosmology where the local volume expansion is driven by the
  domain average density}},  {\em PTEP} {\bf 2019} (2019), no.~7 073E01,
  [\href{http://arxiv.org/abs/1904.09689}{{\tt arXiv:1904.09689}}].

\bibitem{Buchert:1999er}
T.~Buchert, {\it {On average properties of inhomogeneous fluids in general
  relativity. I: Dust cosmologies}},  {\em Gen. Rel. Grav.} {\bf 32} (2000)
  105--125, [\href{http://arxiv.org/abs/gr-qc/9906015}{{\tt gr-qc/9906015}}].

\bibitem{Kenworthy:2019qwq}
W.~D. Kenworthy, D.~Scolnic, and A.~Riess, {\it {The Local Perspective on the
  Hubble Tension: Local Structure Does Not Impact Measurement of the Hubble
  Constant}},  {\em Astrophys. J.} {\bf 875} (2019), no.~2 145,
  [\href{http://arxiv.org/abs/1901.08681}{{\tt arXiv:1901.08681}}].

\bibitem{Buchert:2022zaa}
T.~Buchert, H.~van Elst, and A.~Heinesen, {\it {The averaging problem on the
  past null cone in inhomogeneous dust cosmologies}},
  \href{http://arxiv.org/abs/2202.10798}{{\tt arXiv:2202.10798}}.

\bibitem{Tiwari:2021ikr}
P.~Tiwari, R.~Kothari, and P.~Jain, {\it {Superhorizon Perturbations: A
  Possible Explanation of the Hubble\textendash{}Lema\^\i{}tre Tension and the
  Large-scale Anisotropy of the Universe}},  {\em Astrophys. J. Lett.} {\bf
  924} (2022), no.~2 L36, [\href{http://arxiv.org/abs/2111.02685}{{\tt
  arXiv:2111.02685}}].

\bibitem{Pitrou:2013hga}
C.~Pitrou, X.~Roy, and O.~Umeh, {\it {xPand: An algorithm for perturbing
  homogeneous cosmologies}},  {\em Class. Quant. Grav.} {\bf 30} (2013) 165002,
  [\href{http://arxiv.org/abs/1302.6174}{{\tt arXiv:1302.6174}}].

\bibitem{Brizuela:2008ra}
D.~Brizuela, J.~M. Martin-Garcia, and G.~A. Mena~Marugan, {\it {xPert: Computer
  algebra for metric perturbation theory}},  {\em Gen.Rel.Grav.} {\bf 41}
  (2009) 2415--2431, [\href{http://arxiv.org/abs/0807.0824}{{\tt
  arXiv:0807.0824}}].

\end{thebibliography}
%\include{Fitting_Problem_V12.bbl}

\providecommand{\href}[2]{#2}\begingroup\raggedright\endgroup

\end{document}